\documentclass[iop]{emulateapj}

\usepackage{kotex}
\usepackage{newtxtext,newtxmath}
\usepackage{enumitem}
\usepackage[flushleft]{threeparttable}
\usepackage{esvect}
\usepackage{xcolor}
\usepackage[hidelinks]{hyperref}

\newcommand{\lsim}{{\;\raise0.3ex\hbox{$<$\kern-0.75em\raise-1.1ex\hbox{$\sim$}}\;}}
\newcommand{\gsim}{{\;\raise0.3ex\hbox{$>$\kern-0.75em\raise-1.1ex\hbox{$\sim$}}\;}}

\def\kms{{\rm km s}^{-1}}

\def\kpc{{\rm\thinspace kpc}}

\def\Gyr{{\rm\thinspace Gyr}}

\slugcomment{ }

\shorttitle{Channels producing gas-poor galaxies}
\shortauthors{Jung et al.}

\begin{document}


\title{On the origin of gas-poor galaxies in galaxy clusters using cosmological hydrodynamic simulations}


\author{Seoyoung L. Jung\altaffilmark{1}, Hoseung Choi\altaffilmark{1}, O. Ivy Wong\altaffilmark{2}, Taysun Kimm\altaffilmark{1}, Aeree Chung\altaffilmark{1}, and Sukyoung K. Yi\altaffilmark{1}}
\affil{$^{1}$Department of Astronomy, Yonsei University, 50 Yonsei-ro, Seodaemun-gu, Seoul 03722, Republic
of Korea\\ $^{2}$International Centre for Radio Astronomy Research (ICRAR), University of Western Australia, 35 Stirling Highway, WA 6009, Australia}


\begin{abstract}

The environmental effect is commonly used to explain the excess of gas-poor galaxies in galaxy clusters. Meanwhile, the presence of gas-poor galaxies at cluster outskirts, where galaxies have not spent enough time to feel the cluster environmental effect, hints for the presence of pre-processing. Using cosmological hydrodynamic simulations on 16 clusters, we investigate the mechanisms of gas depletion of galaxies found inside clusters. The gas depletion mechanisms can be categorized into three channels based on where and when they took place. First, 34$\%$ of our galaxies are gas poor before entering clusters (``pre-processing''). They are mainly satellites that have undergone the environmental effect inside group halos. Second, 43$\%$ of the sample became quickly gas deficient in clusters before the first pericentric pass (``fast cluster processing''). Some of them were group satellites that are low in gas at the time of cluster entry compared to the galaxies directly coming from the field. Even the galaxies with large gas fractions take this channel if they fall into massive clusters ($\gsim 10^{14.5}\, \rm M_{\rm \odot}$) or approach cluster centers through radial orbits. Third, 24$\%$ of our sample retain gas even after their first pericentric pass (``slow cluster processing'') as they fall into the less massive clusters and/or have circular orbits.
%that minimize the environmental effect of clusters. 
The relative importance of each channel varies with a cluster's mass, while the exact degree of significance is subject to large uncertainties. Group pre-processing accounts for a third of the total gas depletion; but it also determines the gas fraction of galaxies at their cluster entry which in turn determines whether a galaxy should take the fast or the slow cluster processing. 
%Our results demonstrate that it is important to consider the past history of galaxies before entering current clusters for understanding the origin of gas-poor galaxies in clusters.
\end{abstract}


%% subject header.  Objects should be in the appropriate "individual"
%% headers (e.g. quasars: individual, stars: individual, etc.) with the
%% additional provision that the total number of headers, including each
%% individual object, not exceed six.  The \objectname{} macro, and its
%% alias \object{}, is used to mark each object.  The macro takes the object
%% name as its primary argument.  This name will appear in the paper
%% and serve as the link's anchor in the electronic edition if the name
%% is recognized by the data centers.  The macro also takes an optional
%% argument in parentheses in cases where the data center identification
%% differs from what is to be printed in the paper.

\keywords{methods: numerical -- galaxies: clusters: general -- galaxies: evolution -- galaxies: groups: general}


\section{Introduction} \label{s1}

Earlier observations of external galaxies demonstrated that the physical properties of galaxies depend on their surrounding environment. Galaxies in dense, crowded environments predominantly have early-type morphology (e.g., \citealt{Oemler_1974}; \citealt{Davis_1976}; \citealt{Dressler_1980}; \citealt{Postman_1984}), red color (e.g., \citealt{Pimbblet_2002}; \citealt{Hogg_2003}; \citealt{Balogh_2004}; \citealt{Baldry_2006}), reduced star formation activity (e.g., \citealt{Balogh_1997}; \citealt{Poggianti_1999}; \citealt{Lewis_2002}; \citealt{Kauffmann_2004}; \citealt{Wetzel_2012}), and low neutral hydrogen (HI) content (e.g., \citealt{Haynes_1984}; \citealt{Fabello_2012}; \citealt{Hess_2013}). The environmental effect is commonly used to explain these observed trends. By splitting galaxies into centrals and satellites, researchers found that satellite galaxies are more likely to be red, quenched, and gas poor compared to the central galaxies with the same stellar mass and host halo mass ranges (\citealt{vandenBosch_2008}; \citealt{Weinmann_2009}; \citealt{Wetzel_2012}; \citealt{Catinella_2013}). Considering the hypothesis of \cite{vandenBosch_2008} that galaxies are first born as centrals of their own dark matter halo and then may become satellites as they fall into a larger halo, the distinction in the quenched galaxy fraction between the central and satellite populations appears to be due to the transformation mechanisms that affect satellites predominantly. \cite{Peng_2012} discovered that the dependence of the red galaxy fraction on their local environment vanishes for central galaxies (see also \citealt{Kimm_2009}). Therefore, we can say that the observed excess of red galaxies in dense environments is governed by satellite-specific mechanisms.
%\textcolor{blue}{In this regard, galaxy clusters, where the brightest cluster galaxy (BCG) dominate their satellites in terms of mass, are the best test-beds for studying the environmental effect.}

Galaxy clusters are among the most massive systems in the Universe. In this extreme environment, the host dark matter halo and its brightest cluster galaxy dominate their satellites in terms of mass so that the effect of satellite-specific processes are more prominent in clusters than in less massive systems. In this sense, clusters are the best test-beds for studying the environmental effect.

When focusing on one massive cluster halo, the number of early-type, red, and gas-deficient galaxies is highest at the center of the halo and gradually becomes smaller towards the outskirts (\citealt{Whitmore_1993}; \citealt{Balogh_1999}; \citealt{Gavazzi_2006}; \citealt{Solanes_2001}; \citealt{Gomez_2003}; \citealt{Goto_2003}; \citealt{vonderLinden_2010}; \citealt{Bahe_2013}). By assuming that a radial distance from the cluster center to a galaxy roughly indicates the time since infall to the cluster (e.g., \citealt{Rhee_2017}), the transformation of galactic properties according to their distance from the cluster center can be interpreted as the time evolution of galaxies falling into the cluster. When a galaxy moves from a field environment to a dense environment, it experiences physical processes that are active in dense environments, causing the galaxy to lose its gas. In particular, low-density gas, which is mainly observable in the form of HI emission, is highly fragile to the environmental effect as it is diffuse and typically extended beyond the stellar disk (e.g., \citealt{Hewitt_1983}; \citealt{Fouque_1983}; \citealt{Giovanelli_1983}). The absence of galactic gas leads to the cessation of star formation. If there is no additional supply of cold gas from outside to refuel the star formation, a galaxy soon becomes red.

Indeed, many mechanisms have been proposed to explain how the environment induces the transformation of galactic properties. Due to the high galaxy number density in massive clusters compared to typical regions of the Universe, there is a high probability of gravitational interactions with neighbor galaxies inside clusters. Harassment (\citealt{Moore_1996}) is a fast encounter between galaxies and is common. Mergers induce violent morphology transformation (\citealt{Toomre_1972}) but are assumed to be rare in massive clusters due to high relative speeds. At the same time, a galaxy can interact with the dark matter halo potential, undergoing a tidal stripping of both gas and stars (\citealt{Byrd_1990}; \citealt{Bekki_1999}). X-ray observations covering dense cluster regions have revealed that massive dark matter halos are filled with hot intracluster medium (ICM) (\citealt{Sarazin_1986}; \citealt{Neumann_1999}; \citealt{Pratt_2002}). As a galaxy moves through the ICM, it experiences the ram pressure that strips off the interstellar medium (ISM) and becomes a gas-poor galaxy (\citealt{Gunn&Gott_1972}; \citealt{Quilis_2000}). When the ram pressure is weak and merely able to strip hot surrounding halo gas, the galaxy slowly runs out of the cold gas without additional gas supply from hot gas cooling, i.e., starvation (\citealt{Larson_1980}). Moreover, thermal evaporation due to the hot ICM can lead to the loss of galactic ISM (\citealt{Cowie_1977}). 

Numerous studies have attempted to estimate the gas depletion timescale or star formation quenching timescale to obtain an insight into the physical processes responsible for the environmental effect. Using the Blind Ultra Deep HI Environmental Survey (BUDHIES), \cite{Jaffe_2015} discovered that HI depletion of BUDHIES galaxies takes place before the first pericentric pass. They found a higher fraction of HI detected galaxies among the first infalling galaxies (67$\%$) than the virialized ones (14$\%$) and claim that ram pressure stripping is the key mechanism for gas depletion. 
%\cite{Roediger_2007} performed a suite of 3D hydrodynamic simulations of gas-rich disk galaxies orbiting inside clusters and confirmed that the size of the galactic gas disk closely follows the truncation radius expected from the analytic calculation of ram pressure by \cite{Gunn&Gott_1972}. 
Regarding the star formation quenching timescale, \cite{Mahajan_2011} used phase-space analysis and found that their observed galaxies seem quenched during the first passage through the cluster. Meanwhile, \cite{Wetzel_2013} advocated a ``delay-then-rapid'' quenching model that matches the observed bimodality of the specific star formation rate (sSFR) among SDSS cluster satellites. According to their model, galaxies keep forming stars after they pass their first pericenter. Indeed there is a significant amount of mass increase through star formation activity among their galaxies, between cluster infall and redshift 0. See also \cite{Tollet_2017} who found that the quenching model, in which the star formation activity of infalling galaxies is sustained at least during the first infall, matches the observed stellar mass function in groups and clusters. 

``Pre-processing'' seems to be another channel that produces an excess of gas-poor, quenched populations in clusters. For example, \cite{Wetzel_2013} claimed that the observed difference in the passive galaxy fraction as a function of the host halo mass is only due to the variation in the contribution of the pre-processed galaxies in each halo. Observations on cluster vicinities reported that some galaxies outside the virial radius of the cluster are already red, gas deficient, and quenched despite that they are unlikely to have experienced the environmental effects of the cluster yet (\citealt{Jaffe_2015}; \citealt{Jaffe_2016}). In particular, group-size halos are probable sites of pre-processing (\citealt{Cortese_2006}; \citealt{Just_2015}; \citealt{Bianconi_2018}). Through the hierarchical assembly, group-size dark matter halos dive into massive clusters guiding their associated galaxies, some of which are already in a gas-deficient state, to be members of the clusters. There is growing evidence of a moderate environmental effect in groups, resulting in a higher fraction of early-type, passive, gas-poor galaxies compared to the field environment (\citealt{Postman_1984}; \citealt{Zabludoff_1998}; \citealt{Cybulski_2014}; \citealt{Barsanti_2018}). Therefore, it is important to estimate the significance of pre-processing in producing passive, gas-poor galaxies observed in clusters.

In this study, we use a set of cosmological hydrodynamic simulations (\citealt{Choi_2017}) to explore both pre-processing and the environmental effect inside clusters. The novelty of this simulation is that we have 16 clusters with a variety of halo masses reaching up to $M_{\rm 200} \approx 10^{15}\,M_\odot$ by using the ``zoom-in technique'' from a large cosmological volume (see Section \ref{s2.1} for details). Our goal is to trace back the gas history of galaxies and investigate possible channels that produce gas-poor galaxies found in clusters. More specifically, we try to answer the following questions; (i) when and where did the gas depletion processes take place, and (ii) what is the timescale of gas depletion inside clusters?

This paper is structured as follows. In Section \ref{s2}, we describe our simulations, present the method for measuring the gas properties of galaxies, and define the term “gas-poor galaxy” that is the central concept of this paper. In Section \ref{s3}, we demonstrate what can be inferred from the radial distribution of gas-poor galaxies. Section \ref{s4} provides an in-depth discussion of pre-processed galaxies, galaxies that become gas poor before entering the clusters. This section includes a discussion on how many galaxies are pre-processed and under what conditions is the pre-processing efficient. In Section \ref{s5} we examine galaxies that are depleted of their gas inside the cluster halo. We mainly focus on the timescales of being gas-poor galaxies in clusters and investigate what determines the depletion timescale.  In Section \ref{s6.1}, we briefly discuss the signs of active ram pressure stripping both in groups and clusters. Finally, in Section \ref{s7}, we present three possible scenarios of producing gas-poor galaxies and compare their relative importance.

\section{Methods} \label{s2}

\subsection{Numerical simulation} \label{s2.1}

\begin{figure*}
    \centering
    \includegraphics[width=\textwidth]{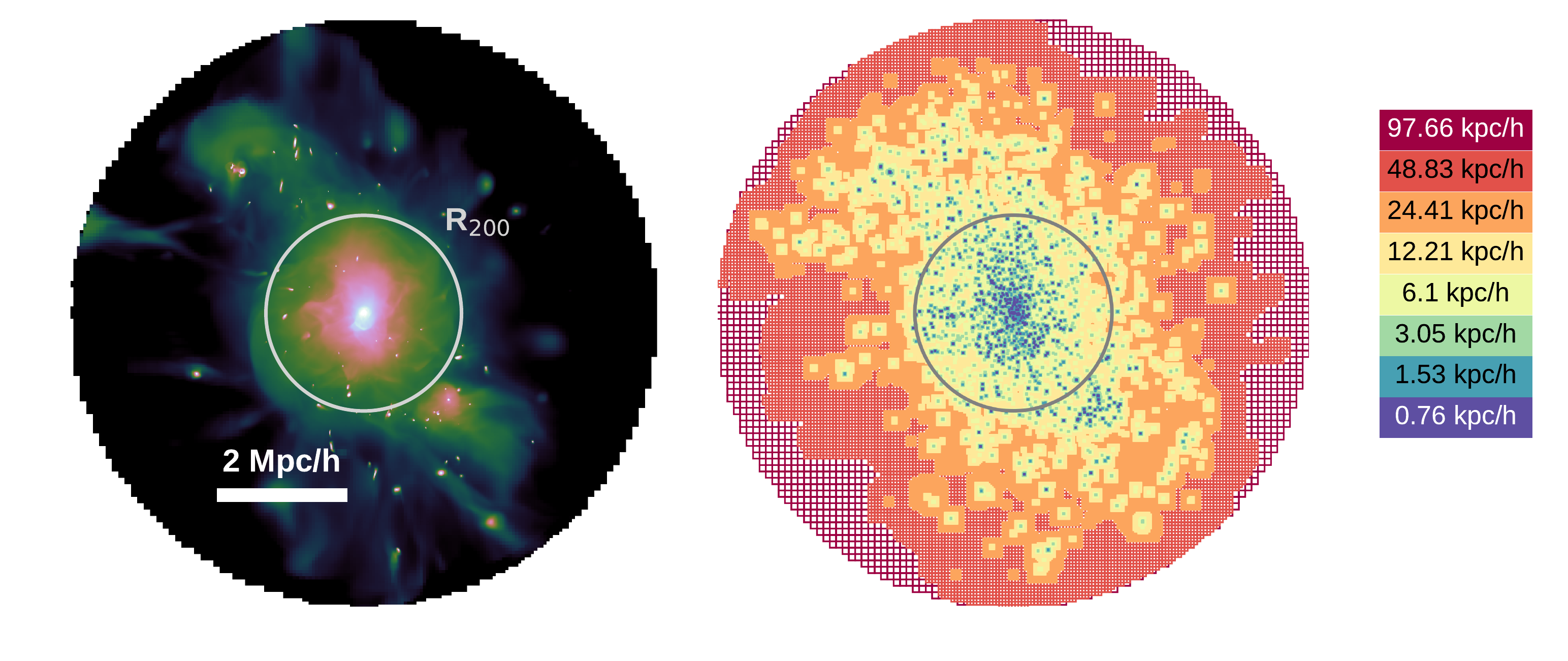}

    \caption{
    Gas density (left) and AMR grid (right) map of gas cells within $3 R_{\rm 200}$ of a cluster with $\log M_{\rm 200}/M_\odot = 14.74$ at redshift 0. Each color-edge square in the grid map represents a gas cell with its spatial resolution that corresponds to the numbers on a box with the same color on the right. The circles at the center present $R_{\rm 200}$ of the cluster.
    }
    \label{fig:f1}
\end{figure*}

We used the Yonsei Zoom-in Cluster Simulation (YZiCS, \citealt{Choi_2017}), which was performed using the adaptive mesh refinement (AMR) code, {\sc ramses} \citep{Teyssier2002}. We adopted a cosmology based on the seven-year Wilkinson Microwave Anisotropy Probe (WMAP) data \citep{Komatsu2011}: a Hubble constant of $H_{\rm 0} = 70.4 \, \kms$ Mpc$^{-1}$, a baryon density of $\Omega_{\rm b} = 0.0456$, a total matter density of $\Omega_{\rm m} = 0.272$, a dark energy density of $\Omega_{\rm \Lambda} = 0.728$, an rms fluctuation amplitude at $8\,h^{-1}\,{\rm Mpc}$ of $\sigma_{\rm 8} = 0.809$, and a spectral index $n = 0.963$. We first ran a dark-matter-only simulation for a $200\, h^{-1} {\rm Mpc}$ cube from which 16 clusters were selected and resimulated with the maximum physical resolution of $0.76 \,h^{-1}\,{\rm kpc}$. Thanks to the large parent volume, our sample contains galaxy clusters of mass $5\times10^{13} < M_{\rm 200}/M_\odot < 10^{15}$. 
The zoom-in simulations cover the volume out to the clustocentric distance to the farthest particle in the initial stage ($z=50$) among all the dark matter particles inside $3 R_{\rm 200}$ of the cluster at $z=0$.
Figure~\ref{fig:f1} shows the projected distributions of gas density and AMR grids within $3\,R_{200}$ of one of the YZiCS clusters with mass $\log M/\rm M_{\odot} \sim 14.74$ at redshift 0. It demonstrates our AMR approach traces the gas structure: all density peaks within $3\,R_{200}$ are resolved at least at $1.5\,\rm h^{-1}kpc$ while low-density regions are resolved at a few kpc/h.
%The high-resolution region of each resimulation contains three times the virial radius of the cluster at $z=0$ to cover various environments from the central region of the cluster to the outskirt. 
The simulation results were saved at a constant interval of a scale factor, $\Delta a_{\rm exp} \approx 0.005$, except for very early epochs, resulting in the total of 187 snapshots.

We adopt the baryon prescriptions of \citet{Dubois2012}. A brief summary is as follows; gas cooling by primodial species (H and He) is modeled assuming collisional ionization equilibrium, while metallcity-dependent cooling is included using \citet{Sutherland1993}. Heating due to the ultraviolet background radiation (\citealt{Haardt1996}) is turned on at $z=10$. Star formation takes place in cells with hydrogen number densities above $n_{\rm 0} = 0.1 \,{cm}^{-3}$ \citep{Rasera2006, Dubois2008} following the Schmidt law $\dot{\rho} = \varepsilon_{\rm \ast}\rho_{\rm g}/t_{\rm \rm ff}$, where $\rho_{\rm \rm g}$ is the gas density, $t_{\rm \rm ff}=\sqrt{3\pi/(32 G \rho_{\rm \rm g})}$ is the local free-fall time, and $\varepsilon_{\rm \ast}$, the star formation efficiency per free-fall time, is set to $2\%$ \citep{Kennicutt1998}. Feedback from supernovae Type II is modeled assuming a Salpeter \citep{Salpeter1955} initial mass function. Seed black holes with a mass of $10^5\, \rm M_{\rm \odot}$ form in dense cells where $\rho_{\rm 0} > 0.1 \,\rm{cm}^{-3}$ and Jean's criterion is violated. Accretion onto black holes is modeled by the Bondi-Hoyle-Lyttleton accretion \citep{Hoyle1939, Bondi1944}. When the gas accretion rate is low, black holes launch bi-polar jets (radio mode), whereas the active galactic nuclei deposit thermal energy isotropically when the accretion rate is high (quasar mode) \citep{Dubois2012}.

\subsection{Galaxy identification}

The galaxies in the simulation were identified using HaloMaker through the AdaptaHOP method \citep{Aubert2004}, with the most massive sub-node mode \citep{Tweed2009} applied for star particles. The mass resolution of star particles is $5\times10^{5}\, \rm M_{\rm \odot}$. A minimum of 2000 particles, i.e., a stellar mass of $10^{9} \, \rm M_{\rm \odot}$ at $z=0$ are required to be identified as a galaxy. Using a galaxy merger tree, we follow their most massive progenitors up to redshift 2. To avoid galaxies with poorly constructed merger trees, we restrict our sample to galaxies linked in more than 100 snapshots among the 187 snapshots we have. The resulting number of galaxies at $z=0$ are 3818. 

\subsection{Gas properties of simulated galaxies}

\begin{figure}
    \centering
    \includegraphics[width=0.9\columnwidth]{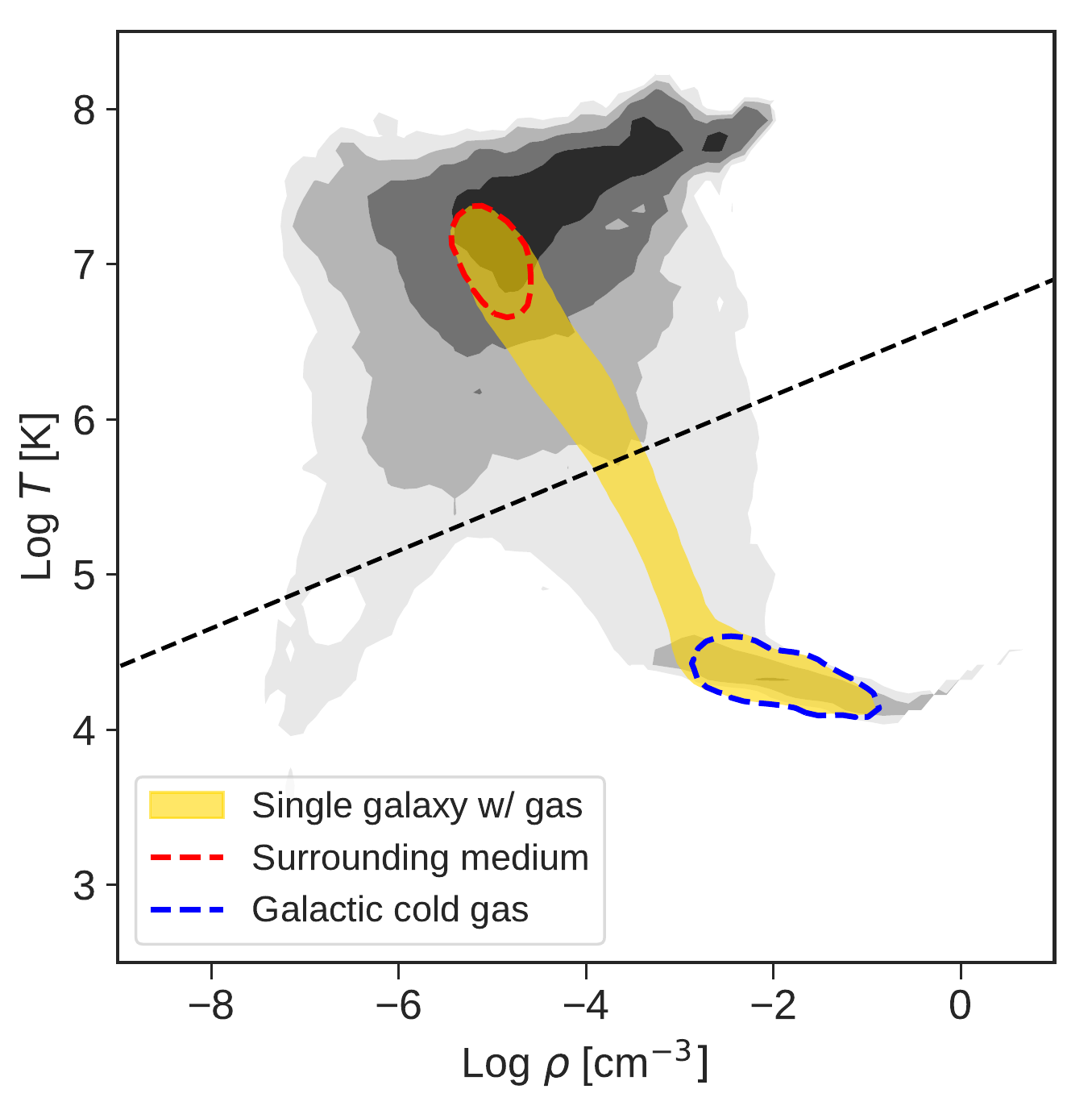}
    \caption{
    Distribution of simulated gas cells in the density-temperature plane, weighted with their mass. The gray shades are the distribution of all the gas cells within 3$\, R_{\rm 200}$ of the cluster, and the yellow shade is 1$\sigma$ distribution of gas cells within and nearby one specific galaxy for example. The black dashed line is from equation (1) of \cite{Torrey_2012}. The red and blue dashed contours are 1$\sigma$ distributions of the ambient medium and galactic gas extracted from the cells in the yellow shade, respectively.}
    \label{fig:f2}
\end{figure}

Due to the Eulerian nature of the AMR technique, gas cells are non-traceable entities throughout time but are redefined in every snapshot instead. Nevertheless, since we are interested in the integrated gas properties of individual galaxies and their surroundings, e.g., the total cold gas mass of a galaxy and surrounding ICM density, the changes over time were tracked by linking the gas properties of galaxies measured at each snapshot with a galaxy merger tree.

The gas cells tend to settle at thermodynamically quasi-stable phases, with an excess of a population at the corresponding density and temperature. In Figure~\ref{fig:f2} the grayscale shades show the mass-weighted 2D histogram of all of the simulated gas cells within 3$\, R_{\rm 200}$ of a cluster at redshift 0. There is an excess of cells at $T \sim 10^{7-8} \, \rm K$ and $\sim 10^{4} \, \rm K$, that correspond to the hot halo gas and galactic gas, respectively. The former phase forms when gas falls into massive dark matter halos ($M_{\rm 200} > 10^{12}\,\rm M_{\rm \odot}$) and is shock heated to the virial temperature, e.g., $T_{200} \sim 10^{7-8} \,\rm K$ for cluster-size halos with $M_{\rm 200}\sim 10^{14}\,\rm M_{\rm \odot}$. This gas is in a quasi-hydrostatic state with their cooling timescale longer than their free-fall timescale and at the same time, feedback from active galactic nuclei also keeps the gas hot. Because our simulations are zoomed in dense regions centered on massive halos, most of the gas cells are in the shock-heated hot phase. At the central region where gas density is sufficiently high, the hot gas cools down and forms the galactic cold phase gas (e.g., \citealt{Rees_1977}; \citealt{White_1978}). At the same time, there is a cold-mode accretion following filaments directly from the outside of the halo that also supplies cold gas to a galaxy, but this channel predominantly takes place in less massive haloes at high redshift (e.g., \citealt{Keres_2005}; \citealt{Dekel_2006}).

Note that our simulation does not resolve extremely cold and dense gas phases that exist in real galaxies, e.g., cold molecular gas. Instead we adopt a polytropic equation of state for the gas cells with their density higher than 0.1$\,\rm cm^{-3}$ with the choice of $\gamma = 4/3$ to prevent gas cells from artificial fragmentation induced from the resolution limit (\citealt{Schaye_2008}). In this case, the minimum gas temperature is determined by the equation of state, and we can find these star-forming gas cells lie on a straight line with a slope of 1/3 on the logarithmic $\rho$ - T plane (note the short tail on the right side of the blue-dashed contour of the galactic gas).

The yellow shade in Figure~\ref{fig:f2} shows the 1$\sigma$ distribution of gas cells within and near one specific galaxy in $\rho$ - T plane. From the shock-heated hot phase in the top left to the cold phase in the bottom right, the gas cells lie in a sequence following a cooling flow. In the following section, we carefully separate the galactic gas and the ambient medium in order to measure the quantities used throughout this paper, e.g., the cold gas content of a galaxy.

\subsubsection{Ambient medium and galactic gas} \label{s2.3.1}

The setting of an appropriate size of the region of interest around a galaxy is an important issue. Typical gas-rich galaxies have their cold gas extended beyond their stellar disk (e.g., \citealt{Hewitt_1983}; \citealt{Fouque_1983}; \citealt{Giovanelli_1983}). Therefore, in order to contain all the cold gas cells belonging to a galaxy, the boundary of the region should be larger than its stellar distribution, but at the same time small enough to avoid contamination of external gas blobs, such as neighboring galaxies. First, we set $R_{\rm 90}$ of the star particles (distance to the top 90$\%$ distant star particle from the galaxy center) as a minimum radius to be probed. Then we increase the radius from $R_{\rm 90}$ to 6$\, R_{\rm 90}$ by 1$\, R_{\rm 90}$ and measure the cold gas mass within each spherical shell. If the cold gas cells reside well within the galactic potential, we expect the cold gas mass within each shell to decrease going outward, but if there is a contribution of external gas blobs the shell mass would increase sharply at a certain point. Quantitatively, the boundary radius is determined when the shell masses match $dm_{\rm i} < \epsilon  dm_{\rm i+1}$ where $dm_{\rm i}$ represents the shell mass at the $i$ th shell. A simple isothermal gas sphere would suggest constant shell mass, i.e., $dm_{\rm i}/dm_{\rm i+1} = 1$. We adopt $\epsilon$ = 1.2 to take into account possible small variations caused by the internal structure of a galaxy. The overall results that are discussed throughout this paper are not significantly affected by choice of $\epsilon$. For instance, the fraction of gas-poor galaxies changes by less than 0.2\% at all snapshots when $\epsilon$ changes from 1.0 to 1.5. If there are no external gas blobs detected within 6$\, R_{\rm 90}$ of the galaxy, the boundary of the region of interest is set to 6$\, R_{\rm 90}$, which is typically greater than 100\,kpc at redshift 0 and roughly comparable with the virial radius of the galaxy. We describe below, how the galactic gas and ambient medium were separated from the chunk of gas within this region.\\

\setlength{\parindent}{0cm}
\textbf{\textit{Galactic cold gas cells}}

\setlength{\parindent}{0.125in}
We first implement a linear-cut in the logarithmic $\rho$ - T plane using equation (1) from \cite{Torrey_2012} (see the black dashed line in Figure~\ref{fig:f2})
\begin{equation}
\log (T / [\rm K]) = 6 + 0.25 \log (\rho / 10^{10} \, [\rm M_{\rm \odot} h^{2} \rm kpc^{-3}]) \,
\end{equation}
taking the low-temperature side. In real galaxies, the ISM can have multiple phases from the hot ionized medium ($\rho \sim 10^{-4 - -2} \, \rm cm^{-3}$, $T \sim 10^{6 - 7} \, \rm K$) to extremely cold and dense molecular clouds ($\rho \sim 10^{2 - 6}\, \rm cm^{-3}$, $T \sim 10 - 20\, \rm K$). In this study we define ``galactic cold gas'' as gas cells roughly in hydrostatic equilibrium; cooled and settled down at $T \sim 10^{4 - 4.5}\, \rm K$. The blue dashed contour in Figure~\ref{fig:f2} shows the distribution of cold gas cells of an example galaxy. Within our simulation box, the cold gas cells have a density range of $\rho > 10^{-3}$ $\rm cm^{-3}$. 
%Note that dense molecular gas could not be reproduced due to the resolution limit. 
With the low-temperature gas cells separated by the linear cut, we perform three sigma clipping on the temperature with the gas cells of densities lower than $0.1\, \rm cm^{-3}$. The gas cells with densities greater than $0.1\, \rm cm^{-3}$, i.e., the star-forming cells in our simulation, are always regarded as cold gas.\\

\setlength{\parindent}{0cm}
\textbf{\textit{Ambient hot gas cells}}
\setlength{\parindent}{0.125in}

To determine the ambient gas cells, e.g., ICM cells, we first impose the same linear cut from \cite{Torrey_2012} but take the high-temperature side. In general, the ambient gas cells have low metallicities compared to the metal-enriched galactic gas cells. Therefore, we perform a three sigma clipping on the metallicity to include the relatively metal-poor cells. The red dashed contour in Figure~\ref{fig:f2} shows the 1$\sigma$ distribution of the selected gas cells around the sample galaxy on the $\rho$ - T plane. Note that these cells have $T \sim 10^{7}\,\rm K$ that correspond to the shock-heated hot phase, as discussed above.

\subsubsection{Definition of gas-poor galaxy} \label{s2.3.2}

\begin{figure}
    \centering
    \includegraphics[width=0.9\columnwidth]{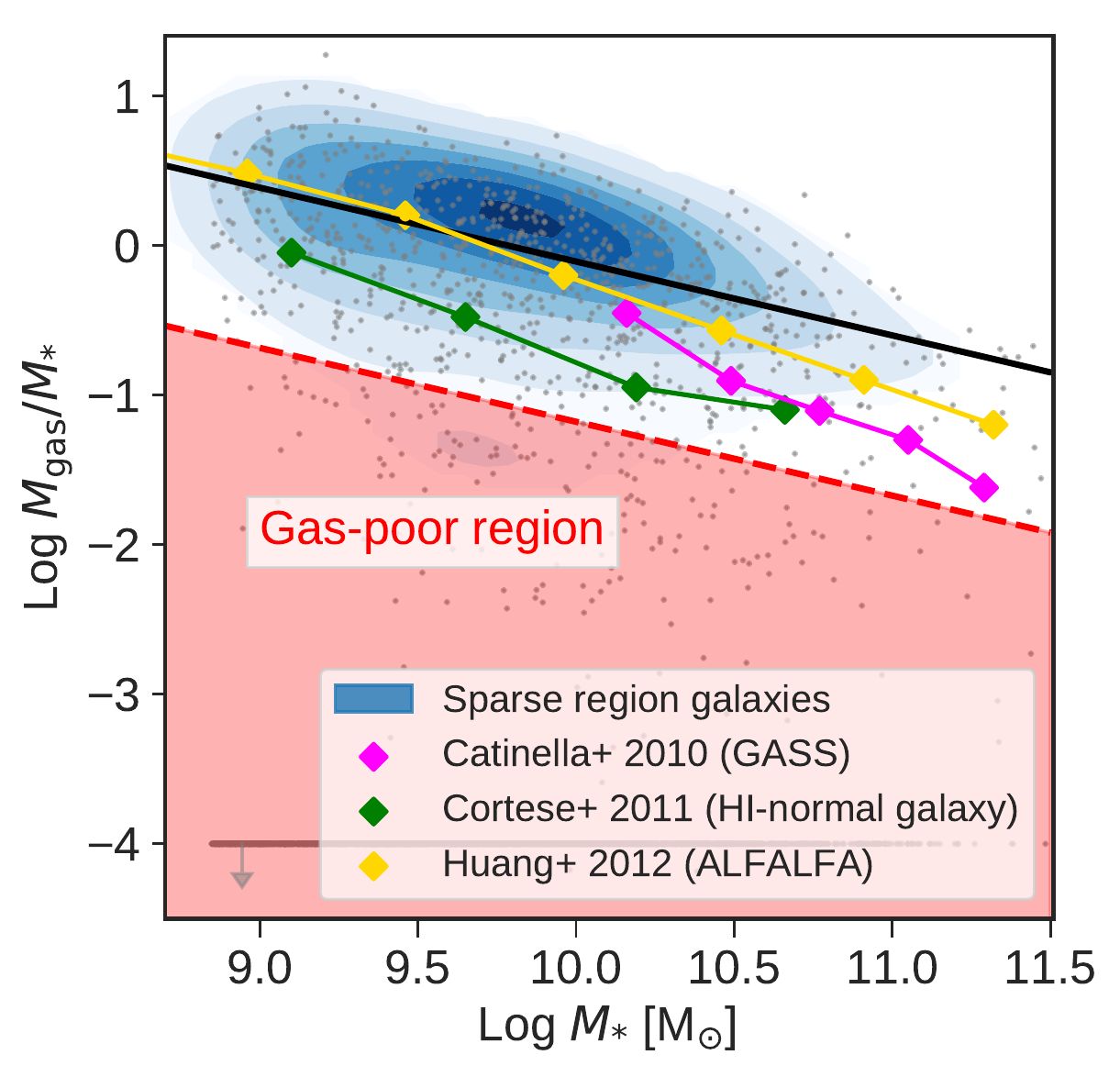}
    \caption{
    The cold gas fraction of YZiCS galaxies measured following the method described in Section \ref{s2.3.1}. The gray dots are the galaxies at redshift 0. The solid black line and blue shades are the linear fit and distribution of the top 20$\%$ low-density region galaxies in terms of a distance to the fifth closest galaxy ($D_{\rm 5}$). Colored lines are the gas fraction from HI observations (\citealt{Catinella_2010}; \citealt{Cortese_2011}; \citealt{Huang_2012}). A galaxy is called ``gas poor'' when it deviates from this gas-rich galaxy sequence more than $\sim 10\%$ (red dashed line, the 2$\sigma$ clipping).
    }  
    \label{fig:f3}
\end{figure}

Large HI surveys of the galaxies in the local Universe have reported an anticorrelation between the gas fraction ($f_{\rm gas} \equiv M_{\rm gas}/M_{\rm *}$) and stellar mass (\citealt{Catinella_2010}; \citealt{Huang_2012}; \citealt{Papastergis_2012}). The scaling relation extends to the galaxies in cluster environments but with an offset toward the lower gas fraction (\citealt{Cortese_2011}). 

In order to compare the gas fraction of YZiCS galaxies with the references, we select the galaxies in low-density regions: top 20$\%$ in terms of the distance to the fifth nearest neighbor at redshift 0. These galaxies are the ones not affected by dense environments, and we confirm that they reveal a similar sequence on the $\log \, M_{\rm *} - \log \, f_{\rm gas}$ plane (blue shades in Figure~\ref{fig:f3}).

In this study, we classify a galaxy as ``gas poor'' when it sufficiently deviates from this gas-rich galaxy sequence toward the lower side. The red dashed line in Figure~\ref{fig:f3} represents a 2$\sigma$ clipping result from the black solid line, the linear fit of the gas-rich sequence, along the y-axis in log scale. Galaxies under this line (the red shaded region) contain roughly 10$\%$ or lower in gas fraction compared to the normal gas-rich galaxies with the same mass, and we call them ``gas poor''. Note that we use a log-scale ($\log \, M_{\rm gas}/M_{\rm *}$) for determining the 2$\sigma$ clipping result because gas fractions of galaxies follow lognormal distributions rather than Gaussian (\citealt{Cortese_2011}).

\section{Radial distribution of gas-poor galaxies} \label{s3}

\begin{figure}
    \centering
    \includegraphics[width=0.9\columnwidth]{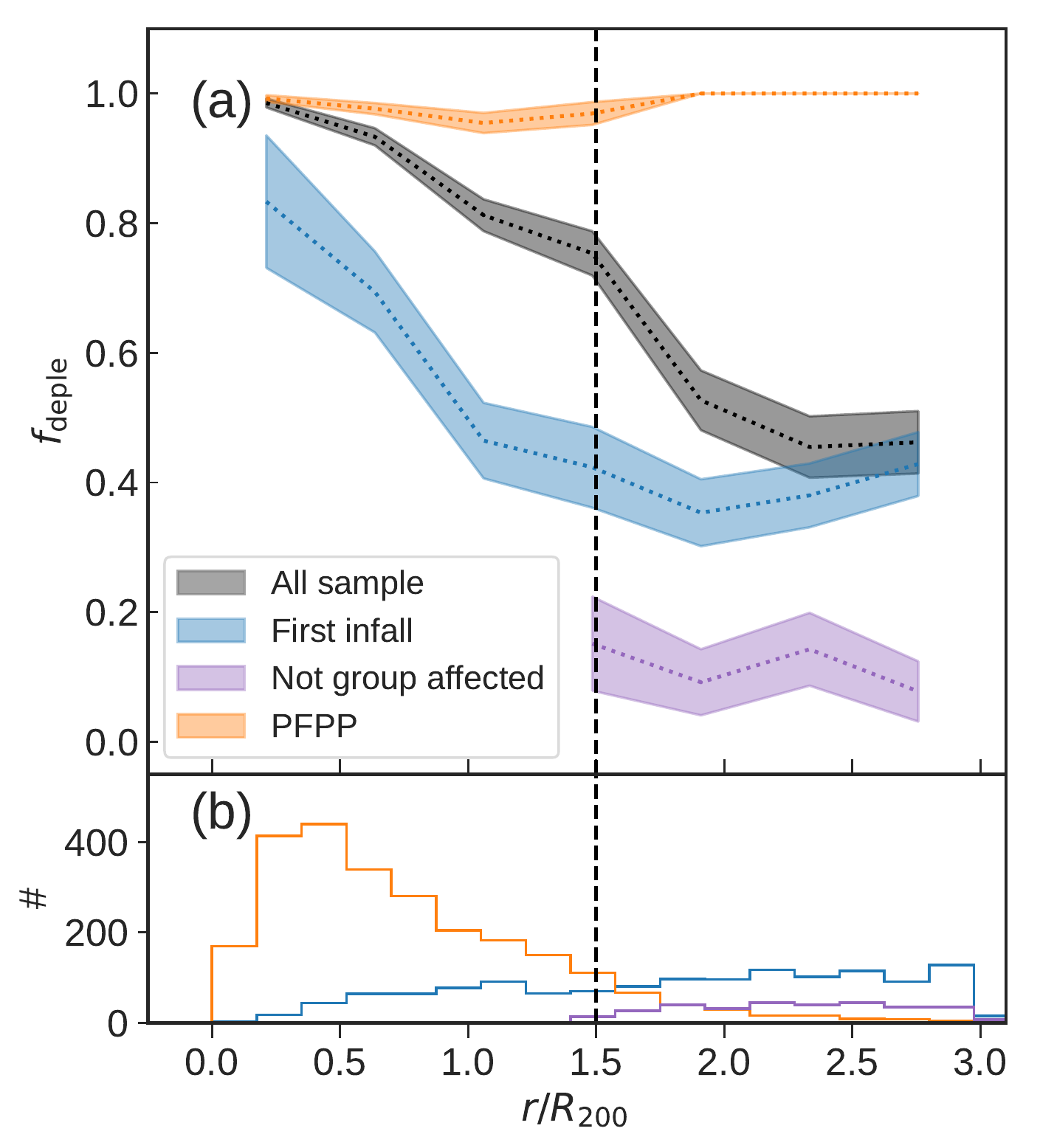}
    \caption{
    (a) The fraction of gas-poor galaxies versus the radial distance from the cluster center of all the galaxies in 16 YZiCS clusters (gray), first infalling galaxies (blue), and ``post first pericentric pass'' (PFPP) galaxies (orange) defined in the text. Among the first infallers, galaxies that have always been central galaxies, i.e., ones that have never affected by the group environmental effect, are shown in purple shade. (b) Radial distribution of each population shown in panel (a). We find evidence of both cluster environmental effect and pre-processing on the galaxies falling into clusters (see text for detailed discussion).
    }  
    \label{fig:f4}
\end{figure}

Observations have reported that gas-poor galaxies are most common near the central region of clusters and their fraction steadily declines further outward, extending far beyond several virial radii of the cluster (e.g., \citealt{vonderLinden_2010}; \citealt{Rasmussen_2012}; \citealt{Wetzel_2012}; \citealt{Barsanti_2018}). The distribution of gas-poor galaxies is the final result of a combination of various gas depletion processes acting in and outside clusters. Using cosmological simulations, \cite{Bahe_2013} suggested three channels that produce the extended radial distribution of gas-poor galaxies, namely pre-processing, overshooting (galaxies moving outward along a highly elongated orbit after their pericentric pass), and direct environmental effect by clusters at large distances. Our sample galaxies also show the steady decline of the fraction of gas-poor galaxies starting from the cluster center out to 3$\, R_{\rm 200}$ (see the gray shade in Figure~\ref{fig:f4}). In this section, we investigate the radial distribution of gas-poor galaxies in order to examine the mechanisms responsible for the gas depletion in our sample galaxies.

Based on whether or not a galaxy passed its first pericenter within a cluster, we separate our sample into ``first-infall'' (blue color) and ``post first pericentric pass'' (hereafter PFPP, orange color) populations. Here, the first pericenter of the orbit is defined as the point where a velocity vector of a galaxy and a direction vector toward the cluster center are perpendicular for the first time. For galaxies approaching cluster centers for the first time, their radial distances from the centers roughly correspond to their time-since-infall (e.g., \citealt{Rhee_2017}). Therefore, by investigating the first-infall population, we can explore the significance of the environmental effect of the cluster as well as pre-processing effects. On the contrary, PFPP galaxies are the ones that have already gone through multiple pericentric passages and sunk down to the central region of the cluster. 

Within 1.5$\, R_{\rm 200}$ of the clusters, there is a radial trend in the fraction of gas-poor galaxies among the first infalling galaxy sample (the blue shade in Figure~\ref{fig:f4} (a)). This strongly suggests that galaxies gradually become gas deficient in their first approach to the cluster center, providing clear evidence for cluster environmental effect. This topic will be discussed further in Section \ref{s5}. 

Meanwhile, at a distance between 1.5 and 3$\, R_{\rm 200}$, the fraction of gas-poor galaxies among the first infalling sample remains nearly constant at $\sim 40\%$. It is important to note that the effect of group pre-processing is visible among the galaxies in this distance range. The purple shade shows the fraction of gas-poor galaxies among the galaxies that have never been a satellite of any halo over their lifetime (these galaxies are subsample of first infalling population). In other words, they have always been a ``central'' galaxy. This fraction is four times lower ($\sim 10\%$) than the fraction of gas-poor galaxies measured with the entire first infalling sample at similar distance ranges ($\sim 40\%$). In other words, galaxies that pass through the group environment are more likely to become depleted in their gas compared to those who have always been a central galaxy. This topic will be discussed in detail in Section \ref{s4}.

We now focus on the origin of the radial trend of the whole sample (grey shade) in the outer region (1.5 through 3 $R_{\rm 200}$). To begin with, the radial trend of the fraction of gas-poor galaxies among the first infalling galaxies (blue shade) is notably weakened at a distance larger than 1.5$\, R_{\rm 200}$. This implies that the cluster environmental effect is not strong until a galaxy falls inside a cluster (see also \citealt{Wetzel_2012}).
%In this case, the PFPP galaxies are mainly responsible for the presence of the radial slope visible in the gray shade. 
On the other hand, PFPP galaxies are almost completely gas poor regardless of their distance from the cluster center (see orange shade). 
%Therefore, the number of PFPP galaxies affects the fraction of gas-poor galaxies at a given distance. 
Note that the orange histogram in panel (b) shows the radial distribution of PFPP galaxies. The majority of the PFPP galaxies ($\sim 91\%$) reside within 1.5$\, R_{\rm 200}$ of the cluster. In contrast, some PFPP galaxies with elongated orbits easily reach beyond virial radius. Quantitatively, $\sim 40\%$ of the galaxies between 1.5 and 2$\, R_{\rm 200}$ are PFPP and their contribution decreases to $\sim 14\%$ between 2 and 2.5$\, R_{\rm 200}$. PFPP galaxies hardly reach distances larger than 2.5$\, R_{\rm 200}$ (\citealt{Mamon_2004}). As a consequence, even though there is no radial trend among the first infalling galaxies beyond 1.5$\, R_{\rm 200}$, the decreased population of PFPP galaxies with increasing radial distance from the center results in decline in the gas-poor galaxy fraction of the whole sample (grey shade).

%If there was an environmental effect induced by the extended cluster and it had a critical impact on the production of gas-poor galaxies, its effect would have been stronger closer to the cluster center causing the fraction of gas-poor galaxies to show a radial trend even at distances larger than 1.5$\,R_{\rm 200}$. However, the purple shade in Figure~\ref{fig:f4} remains constant at all distance ranges. We conclude that the extended cluster effect does not play an important role in producing gas-poor galaxies outside 1.5$\, R_{\rm 200}$ . %The idea that galaxies do not feel the environmental effect until they reach 1.5$\, R_{\rm 200}$ of the clusters This seems to be in conflict with the findings from X-ray observations that show hot ICM reaching out to several virial radii of clusters (\citealt{Frenk_1999}). 
%This inconsistency arose because we monitor the cold gas components of galaxies while mild environmental effects caused by the extended halo preferentially affects the hot halo gas that is easier to be stripped than the cold gas (\citealt{McCarthy_2008}). This can be explained by the fact that we are here monitoring cold gas while X-ray detects hot gas. In fact, our simulation shows the radial profiles of the hot ICM that reach out to XX R200, being consistent with the X-ray observations.

In summary, a hint of pre-processing seems clear and cluster processing is predominantly at work inside $\sim R_{\rm 200}$. The overall radial trend of gas-poor galaxy fraction is a result of the combination of pre-processing and cluster processing effects. Based on the discussions made in this section, we explore each process in detail in the following Sections \ref{s4} and \ref{s5}.

%In summary, by investigating the fraction of gas-poor galaxies as a function of the distance from the cluster centers, we discovered evidence of both cluster environmental effects and pre-processing on the galaxies falling into clusters. A galaxy feels the environmental effect of clusters starting from 1.5$\, R_{\rm 200}$, and in most cases, only a single crossing to the center is sufficient to become gas-deficient. At distances far from the center ($> 1.5\, R_{\rm 200}$), the cluster halo effect diminishes but group pre-processing exists. The extended radial slope on the gas-poor galaxy fraction at this distance range is mainly caused by the distribution of PFPP galaxies. In this study, we did not find the direct environmental effect by clusters producing cold gas-poor galaxies. 

\section{Gas depletion outside cluster halo : pre-processing} \label{s4}

In this section, we investigate the significance of pre-processing effects in detail.
%In the previous section, we demonstrated that pre-processing acting on YZiCS galaxies exists, increasing the fraction of gas-poor galaxies at the outside of the clusters. In this paper, 
We define ``pre-processed galaxy'' as a galaxy that becomes ``gas poor'' (see Section 2.3.2) in at least one snapshot prior to the cluster entry. Therefore, the pre-processing indicates the integrated effect of unspecified mechanisms that induce gas depletion before the cluster entry.

Pre-processed galaxies may be replenished with cold gas afterwards, but we find replenishment to be rare in our simulation. This can be attributed to the fact that virialized hot gas cools inefficiently in the massive halos. In the case of satellites, the ram pressure stripping continuously prevents the gas from collapsing onto the galaxies. A small number of central galaxies are occasionally rejuvenated by merging with gas-rich satellites but central galaxies account for a very small fraction of our pre-processed sample ($\sim2\%;$ 55/3044). In addition, even when a galaxy is rejuvenated, we find that it normally lasts less than a $\Gyr$, and the majority of the rejuvenated galaxies return to the gas-poor state before they reach $1.5R_{200}$ of the cluster halo.

%In this section, we first discuss the method of obtaining a proper parameter that reflects a galaxy's group environment history before entering clusters. Then we investigate the circumstances for a galaxy to be pre-processed in order to obtain an insight into the physical process responsible for the gas depletion outside clusters. And then, we compare the importance of pre-processing in producing a gas-poor galaxy population in our 16 YZiCS clusters.

\subsection{Which galaxies arrive gas deficient?}

\begin{figure}
    \centering
    \includegraphics[width=0.9\columnwidth]{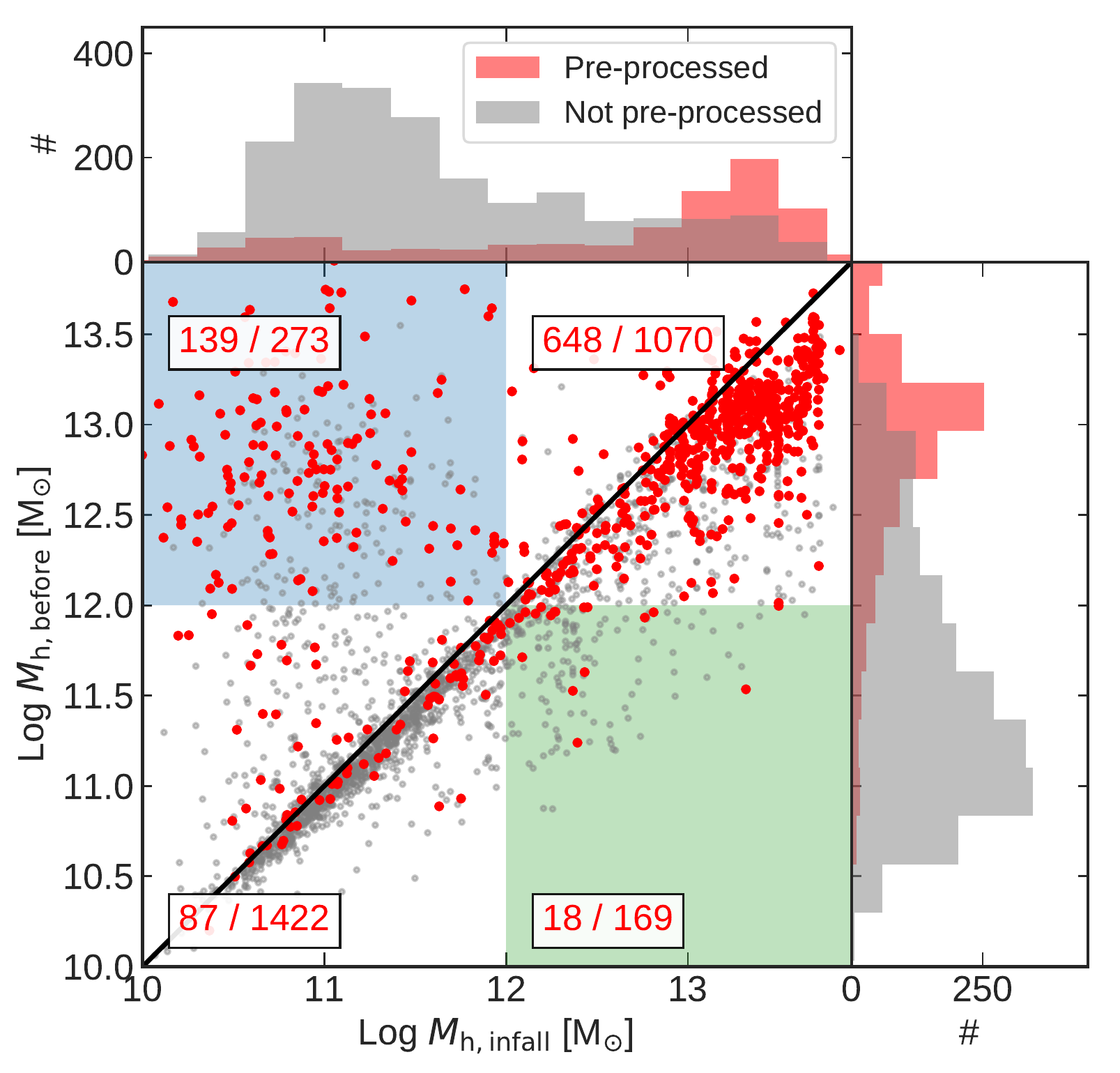}
    \caption{
    Comparison between the host mass at the time of the cluster infall ($M_{\rm h,infall}$) and arithmetic mean value of the host halo masses since $z\sim2$ until the galaxy reaches the cluster ($M_{\rm h, before}$). The red and gray points show the pre-processed galaxies and those entering clusters with gas, respectively. The histograms in the top and right panels show the distribution of $M_{\rm h,infall}$ and $M_{\rm h, before}$ of each population. The blue colored area is where $M_{\rm h,infall}$ underestimates the cumulative environmental effect on galaxies before the cluster infall ($M_{\rm h,infall}<M_{\rm h, before}$) and the green area is the opposite ($M_{\rm h,infall}>M_{\rm h, before}$). The numbers in the boxes on each quadrant are the fractions of the pre-processed galaxy within each quadrant. The pre-processed galaxy fraction better correlates with $M_{\rm h, before}$ than $M_{\rm h,infall}$. 
    %We will use $M_{\rm h, before}$ as an indicator of the environmental effect before the cluster infall throughout this paper.
    }  
    \label{fig:f5}
\end{figure}

\begin{figure*}
    \centering
    \includegraphics[width=0.9\textwidth]{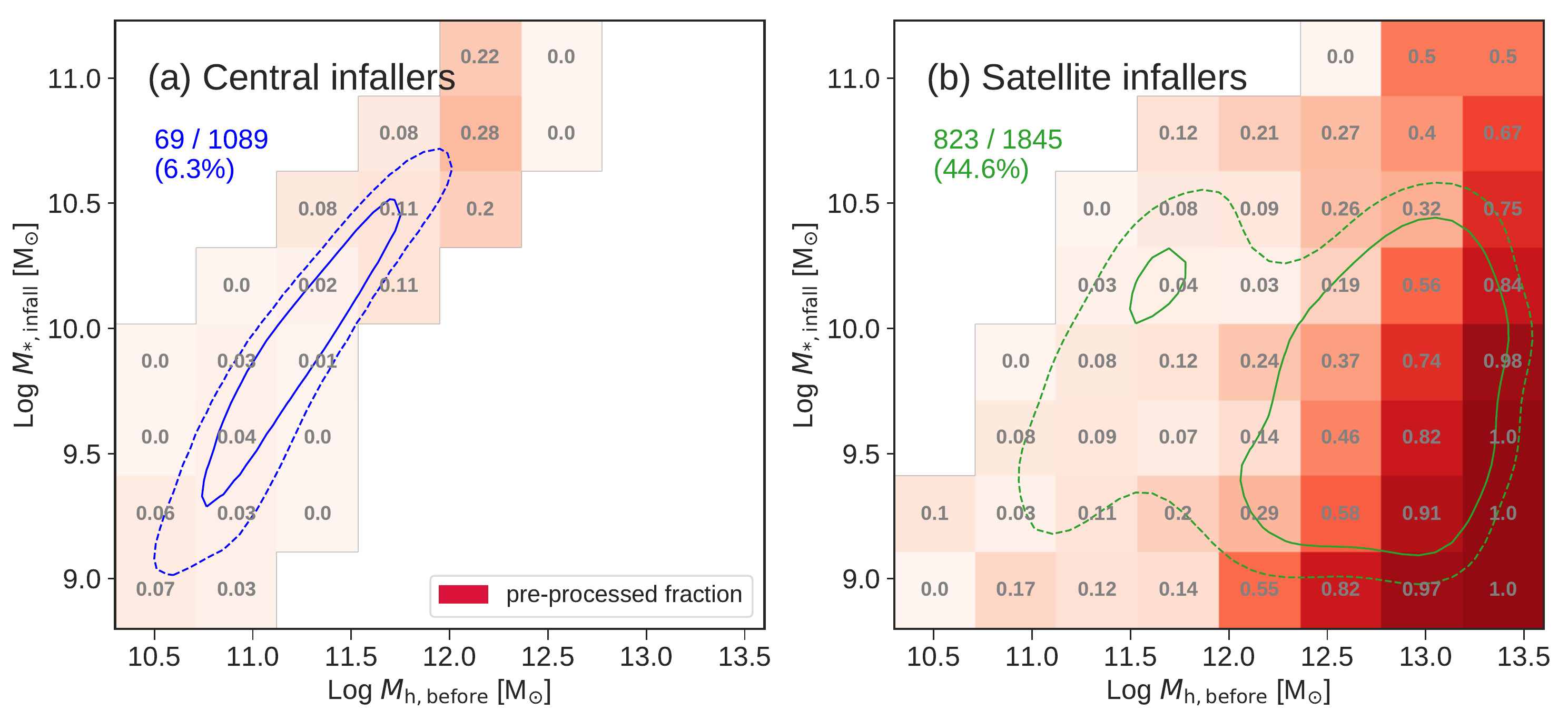}
    \caption{
    Fraction of pre-processed galaxies versus $M_{\rm h, before}$ and $M_{\rm *, infall}$. Each panel corresponds to the (a) central and (b) satellite infallers, respectively. The contours show 1$\sigma$ (dashed line) and 0.5$\sigma$ (solid line) distribution of the galaxies. The numbers on the top left of each panel show the number of the pre-processed galaxy over the total number of galaxies entering the clusters. The satellite infallers have seven times higher pre-processed fractions compared with the central infallers. The red color and numbers on each bin are the fraction of the pre-processed galaxy within the bin. Pre-processing is extremely efficient when the mass ratio between the host halo and satellite galaxy ($M_{\rm h, before}$/$M_{\rm *, infall}$) is high.
    }
    \label{fig:f6}
\end{figure*}

The significance of pre-processing has been a controversial issue. Assuming that dark matter halos of mass greater than $10^{11.5}\,{\rm M_\odot}$ host a galaxy, \cite{Berrier_2009} found that the majority of the infalling galaxies ($\sim 70\%$) arrive at clusters directly from the field, without any neighbor galaxy within their halo. For clusters with a mass range of $10^{14 - 14.76} h^{-1} \rm M_{\rm \odot}$, they argued that pre-processing is uncommon. On the contrary, using the {\sc galform} semi-analytic model of galaxy formation, \cite{McGee_2009} found that a significant fraction of galaxies enters clusters as group-associated galaxies. Their sample covered massive clusters up to $10^{15.3} h^{-1} \rm M_{\rm \odot}$, and they discovered that galaxies are more likely to be pre-processed in more massive clusters. 

Note that both studies define the pre-processed population as galaxies associated with massive halos at the time they arrive at the clusters ($M_{\rm h,infall}$>10$^{13}\,h^{-1}\rm M_{\rm \odot}$). It is true that galaxies in more massive groups have a higher chance of being in a gas-deficient state, but this hypothesis may misestimate the fraction of pre-processed galaxies in the following circumstances. In the real Universe, not all galaxies within massive halos are gas deficient, and at the same time, some gas-poor galaxies are not associated with a group at the moment of the cluster infall. The former occurs when a satellite does not spend sufficient time to experience the group environment. On the contrary, the latter occurs when a satellite escapes its group halo (or appears to have escaped due to its elongated orbit) after spending sufficient time to suffer the gas depletion. Therefore, instead of using the host mass at the time of the cluster infall ($M_{\rm h,infall}$), another parameter should be introduced that better reflects the history of the environmental effect.

In Figure~\ref{fig:f5}, we compare two parameters that estimate the host halo mass of infalling galaxies; $M_{\rm h,infall}$ introduced above and the arithmetic mean value of the host halo masses since $z\sim2$ until the galaxy reaches the cluster ($M_{\rm h, before}$). In the central panel, points represent individual galaxies that have entered clusters. The red points are the pre-processed galaxies and gray points are all the others. The top and right panels show the distribution of $M_{\rm h,infall}$ and $M_{\rm h, before}$ of the two populations. The colored regions designate the areas where $M_{\rm h,infall}$ severely underestimates $M_{\rm h, before}$ (upper left with blue shading) and where $M_{\rm h,infall}$ severely overestimates $M_{\rm h, before}$ (lower right, green). For visual convenience, we introduced a reference line of $10^{12} \, \rm M_\odot$ in this figure. 
%set the characteristic mass of group environmental effect at $10^{12} \, \rm M_{\rm \odot}$. %However, note that this setting is only used to divide regions on the $M_{\rm h, before}$ - $M_{\rm h,infall}$ plane and our definition of the pre-processed galaxy is irrelevant to the characteristic mass.

The numbers in the box in each quadrant show the fraction of pre-processed galaxies. In the top left section (blue), the pre-processed fraction is as high as 50\% despite that their $M_{\rm h,infall}$ is low. This highlights the shortcoming of the use of $M_{\rm h,infall}$ as a criterion for pre-processing, as was the case in some previous studies.
%Note that the pre-processed galaxy fraction in the underestimation region (blue shading, $\sim 51\%$) is as high as the fraction among the galaxies with high $M_{\rm h,infall}$ and high $M_{\rm h, before}$ ($\sim 61\%$). The galaxies in this region are the ones that once resided in massive group halos that are capable of inducing pre-processing but at some point escaped (or appearing to have escaped) the group. 
On the other hand, only 10\% of the galaxies are ``pre-processed'' in the bottom right section (green). Though they are in a relatively massive group at the cluster infall moment, only a small fraction of galaxies appear to be pre-processed mostly because they have not spent enough time in the current group halo yet.
%At the same time, the fraction of pre-processed galaxies in the overestimation region (green shading, $\sim 11\%$) is as low as the fraction in the low $M_{\rm h,infall}$ and low $M_{\rm h, before}$ region even though the galaxies in that area have high $M_{\rm h,infall}$. These galaxies joined massive groups right before the cluster infall, so they did not have sufficient time to feel the environmental effect within the groups. 
The histograms in the two panels on the top and right sum it up: $M_{\rm h, before}$ separates pre-processed and non-pre-processed galaxies better, and thus we have decided to use it in our analysis.
%Considering this discussion, we will make a use of $M_{\rm h, before}$ as an indicator of the group environmental effect before entering clusters throughout this paper.

We will now investigate the conditions for which pre-processing becomes most efficient. To see whether the pre-processing mechanism found in our simulation is due to the environmental effect, we first separate our YZiCS sample galaxies into two groups: namely, the central and satellite infallers. At each snapshot, we determine the central and satellite galaxies by comparing their stellar mass at that moment, so that the central galaxies are always more massive than the companion satellites within their host dark matter halo. Note that the location of galaxies within group halos are not considered when classifying the central and satellites. The sample is then separated into ``central infallers'' and ``satellite infallers''. ``Central infallers'' are the galaxies that have always been the central galaxy of their groups before themselves becoming satellites of clusters. The ``satellite infallers'' are the remaining; the galaxies that have belonged to groups as satellite galaxies before the cluster infall.

Figure~\ref{fig:f6} shows the pre-processed galaxy fraction (red color and white numbers on each bin) depending on the $M_{\rm h, before}$ and $M_{\rm *, infall}$, where $M_{\rm *, infall}$ is the stellar mass at the time of the infall. The sample galaxies are separated into (a) central infallers and (b) satellite infallers. The numbers on the top left of each panel show the number of pre-processed galaxies divided by the total number of galaxies that entered clusters. The satellite infallers have an approximately seven times larger pre-processed fraction than that of the central infallers. Among the 823 satellite infallers, only 30 (3.6$\%$) became gas poor as central galaxies and then later joined a larger group, before the cluster entry. Rest of them deplete their gas as group satellites. 

Pre-processing seems significant considering that 45\% of satellite infallers become gas poor even before cluster entry. It is more significant on satellites (45\%) than in centrals (6\%) and on a smaller satellite in a more massive group. This seems natural because galaxies in more massive halos undergo stronger ram pressure as they normally go through a denser intra-group medium with a higher orbital velocity. At the same time, considering that gas stripping through ram pressure is achieved through a competition between ram pressure and gravitational restoring force, lower mass galaxies lose their gas more efficiently under the equal intensity of ram pressure. The importance and caveats on this issue will be discussed more in Section \ref{s6}.
%Therefore, the excess of pre-processed galaxies among satellite galaxies is still valid, supporting the concept that the satellite-specific environmental effect is mainly responsible for the pre-processing. In addition, among the satellite infallers, the pre-processing is most efficient when the mass ratio between the host halo and galaxy ($M_{\rm h, before}$/$M_{\rm *, infall}$) is high.
%In summary, the mechanism producing pre-processed galaxies is most efficient for (i) low stellar mass, (ii) satellite galaxies, and (iii) in massive group halos. As introduced in Section \ref{s1}, our findings agree with the observations and demonstrate that the environmental effect in group size-halos that specifically affecting satellite galaxies, may be responsible for the pre-processing. We suggest ram pressure stripping acting in groups as a primary mechanism that induces pre-processing outside clusters. Galaxies in more massive halos undergo stronger ram pressure as they normally go through a denser intragroup medium with higher orbital velocity. At the same time, as the gas stripping is achieved through a competition between ram pressure and gravitational restoring force, lower mass galaxies lose their gas more efficiently than massive galaxies under the equal intensity of ram pressure. The importance of ram pressure in group-size halos will be discussed more in Section \ref{s6.1}.

\begin{figure}
    \centering
    \includegraphics[width=0.9\columnwidth]{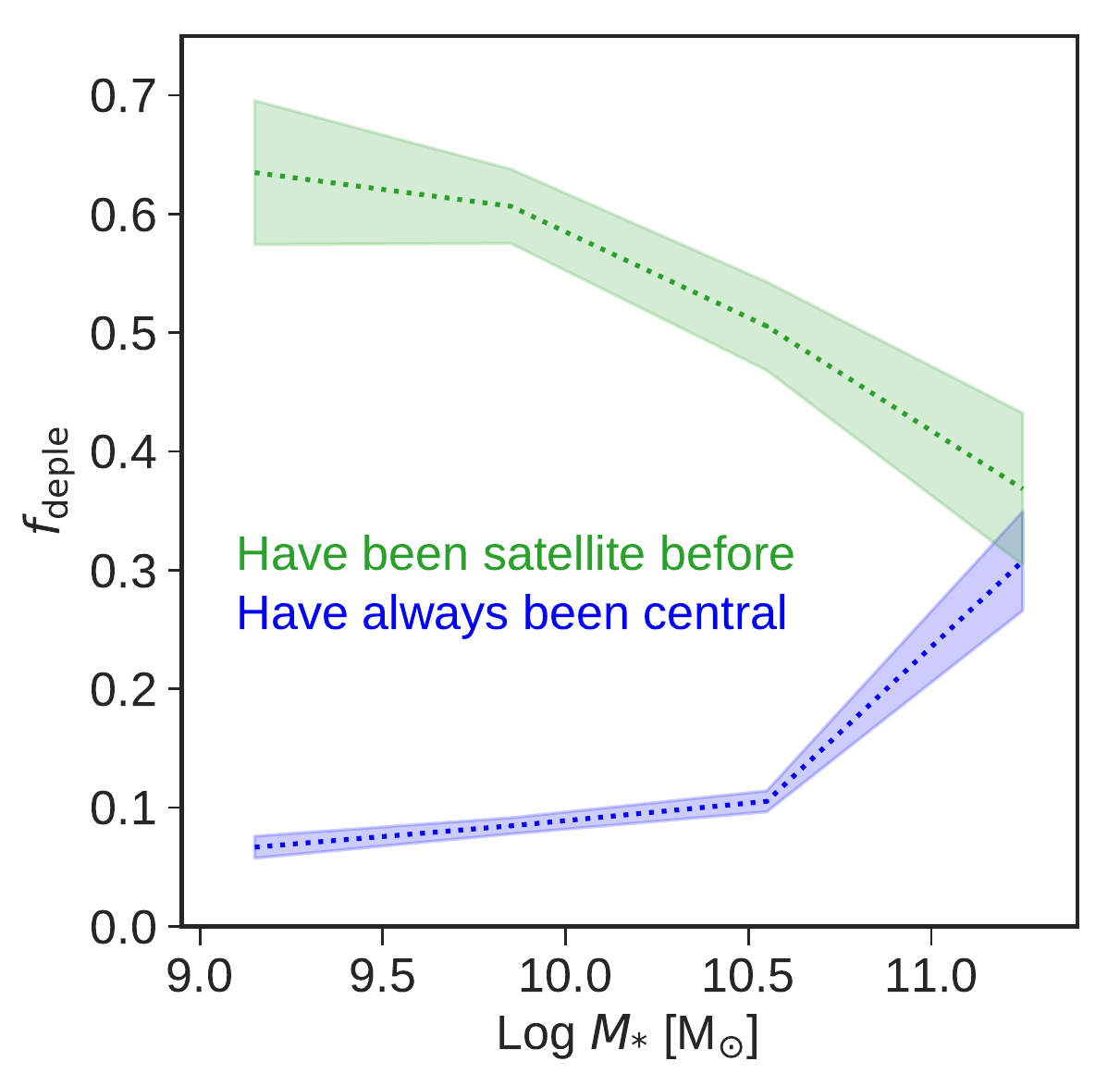}
    \caption{
    The fraction of gas-poor galaxies as a function of stellar mass. Only the galaxies that have not yet fallen into clusters until redshift 0 are presented. Of the galaxies that have always been central galaxies (blue), more massive galaxies are more likely to be gas poor. This trend agrees with the concept of ``mass quenching''. On the contrary, for galaxies that have been satellites (green), the environmental effect is extremely efficient so that lower mass galaxies are more likely to be gas poor.
    }
    \label{fig:f7}
\end{figure}

This result of more massive galaxies having a higher chance of containing cold gas is seemingly in conflict with the concept of ``mass quenching'' (e.g., \citealt{Kauffmann_2003}; \citealt{Baldry_2004}). However, in the following discussion, we confirm that this is not the case. Figure~\ref{fig:f7} shows the fraction of gas-poor galaxies versus their stellar mass for galaxies that have not yet fallen into clusters until redshift 0. The green and blue lines represent galaxies that have been satellites of larger groups and galaxies that have always been central galaxies, respectively. By separating these two populations, we find opposite trends. As discussed above, satellite galaxies more efficiently become gas deficient when their stellar mass is small. Conversely, central galaxies show increased gas-poor galaxy fractions with increasing stellar mass, which agrees with the expectation from mass quenching. That is, we verify that there is an indication of the mass effect among our YZiCS galaxies, but the environmental effect that is more efficient for low mass galaxies overturns the trend. Indeed, using SDSS group galaxies, \cite{Weinmann_2009} and \cite{Wetzel_2012} showed that, at the low stellar mass, satellite galaxies are predominantly redder and more likely to be star formation quenched, compared with the central galaxies (see also \cite{Catinella_2013}). Besides, the distinction between central and satellite galaxies gets weaker when considering massive galaxies. Their results are all consistent with our simulation.

\subsection{Pre-processed fraction depending on cluster mass}

\begin{figure}
    \centering
    \includegraphics[width=0.9\columnwidth]{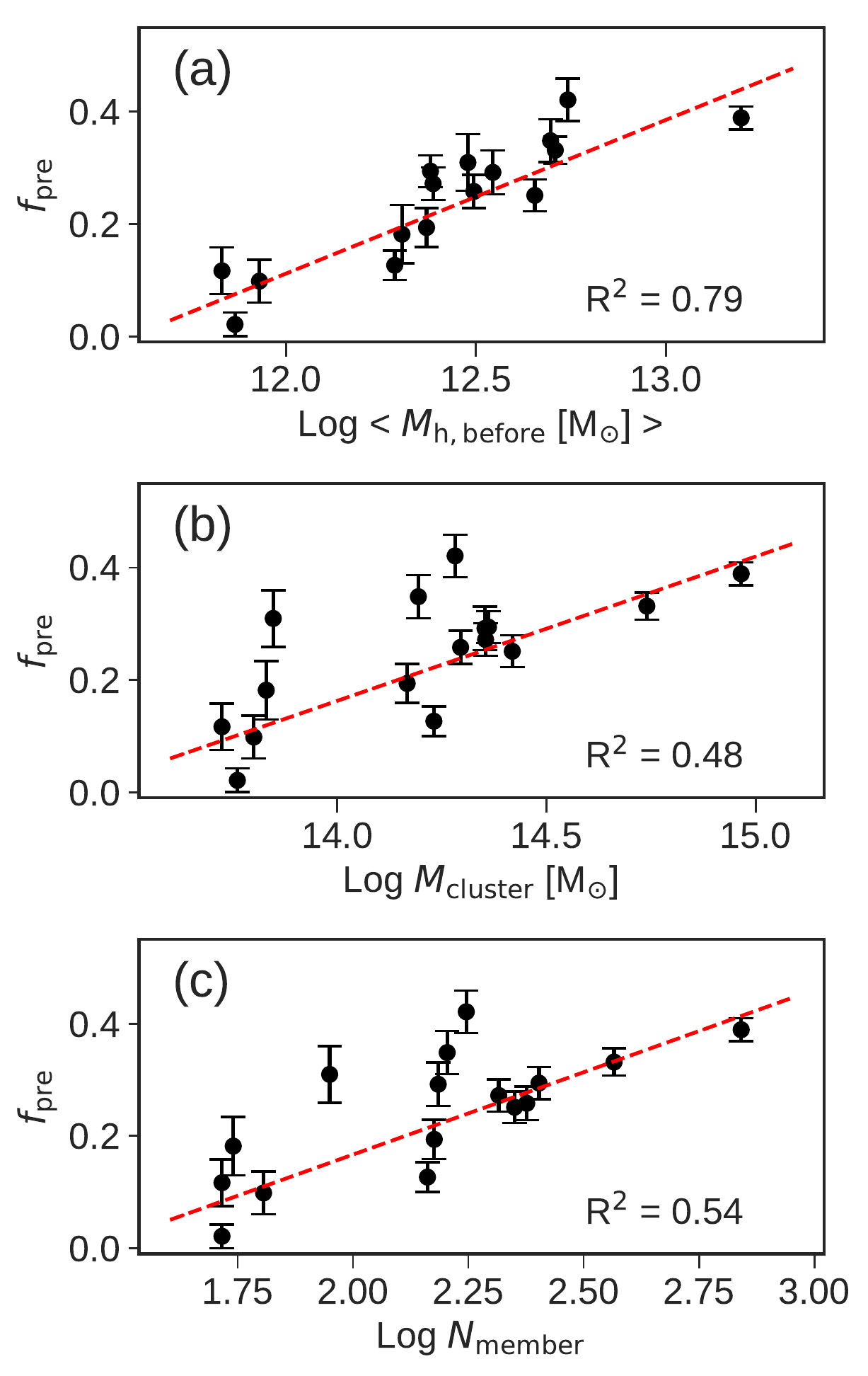}
    \caption{
    Comparison of the fraction of pre-processed galaxies of YZiCS cluster sample with different (a) average of $M_{\rm h, before}$ values of infalling galaxies, (b) cluster halo mass, and (c) the number of cluster member galaxies. Error bars are the standard error of the mean of a binomial distribution. Red dashed line in each panel shows the linear fit and $R^{2}$ presented on the bottom right of the panel is the Pearson coefficient of determination of the fitting.}  
    \label{fig:f8}
\end{figure}

Now we examine the pre-processed fraction of each cluster separately to see whether there are any indications of a correlation between the pre-processed fraction and cluster properties. As shown in the previous section, the amount of pre-processed galaxies is linked with how many galaxies underwent the environmental effect in the massive group-size halos before the cluster infall. For clusters that accreted larger number of massive groups and their constituent galaxies, a higher pre-processed galaxy fraction is expected (see panel (a) of Figure~\ref{fig:f8}). For each cluster we measure the arithmetic average of $M_{\rm h, before}$ of cluster member galaxies ($\langle M_{\rm h, before} \rangle$) to estimate the representative degree of the environmental effect before falling into the clusters. The fraction of pre-processed galaxies linearly increases with increasing $\langle M_{\rm h, before} \rangle$ (see red dashed line that is the linear fit).

\begin{figure}
    \centering
    \includegraphics[width=0.9\columnwidth]{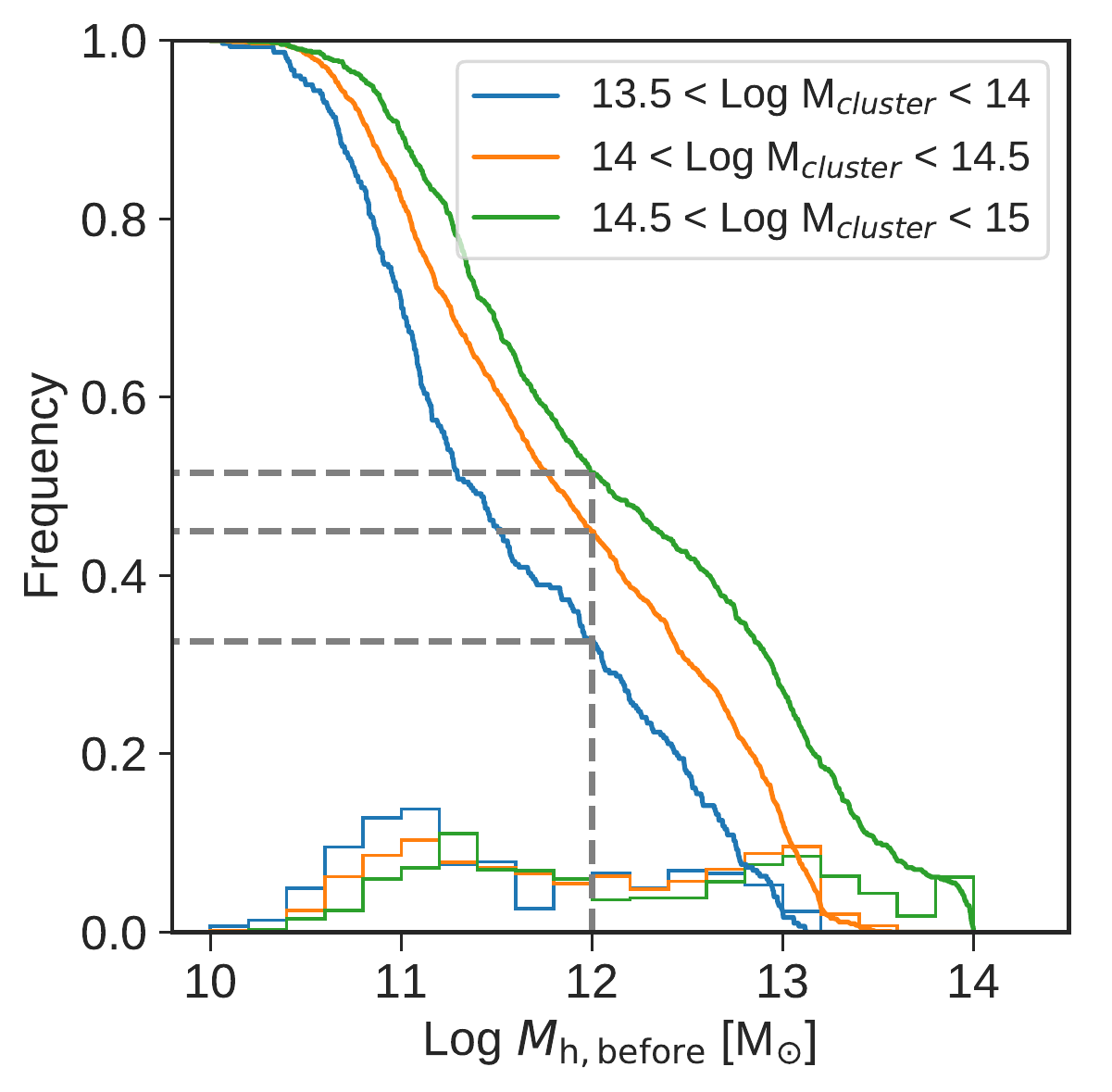}
    \caption{
    Histogram and a cumulative frequency of $M_{\rm h, before}$ for the given range of the cluster mass. More massive clusters accrete larger fraction of galaxies that are affected by massive group halos.
    }
    \label{fig:f9}
\end{figure}

Meanwhile, using Figure~\ref{fig:f9}, we discover that more massive clusters have a higher chance of accreting massive group halos than less massive clusters (\citealt{McGee_2009}). By separating our clusters into three different mass ranges, we compare the $M_{\rm h, before}$ values of their member galaxies. Each solid line represents a cumulative fraction of galaxies with higher $M_{\rm h, before}$ than a given value. For instance, the fraction of galaxies with $M_{\rm h, before}$ larger than $10^{12}\, \rm M_{\rm \odot}$ increases from $\sim 32\%$ to $\sim 45\%$ and $\sim 51\%$ as the cluster mass range increases (see the gray dashed guiding lines). As a consequence, the pre-processed fraction also increases with increasing cluster mass (Figure~\ref{fig:f8}, panel (b)) and the number of cluster member galaxies \footnote{The cluster member galaxies are defined as galaxies that are more massive than $> 10^{9}\, \rm M_{\rm \odot}$ and reside within 1.5$\, R_{\rm 200}$ of the cluster.} as well (Figure~\ref{fig:f8}, panel (c)). However, comparing $R^{2}$, i.e., the Pearson coefficient of determination of the linear fit, we find larger scatter in panel (b) and (c) compared to (a).

We interpret these results as follows: during the hierarchical build-up, more massive clusters are more likely to reside and grow in more crowded regions with a higher chance of accreting massive, rich groups. Within massive group-size halos, the environmental effect on their satellites is strong enough to make a good fraction of satellite galaxies gas deficient, even before reaching main clusters. As a result, the contribution of pre-processing on building a gas-poor galaxy population in clusters becomes larger in more massive clusters. Yet, the scatter in the pre-processed fraction among clusters of similar mass and member counts implies that individual galaxy accretion histories of clusters are a fundamental factor in determining the pre-processed galaxy fraction.

\subsection{Summary on pre-processing}

%We explored the conditions whereby the pre-processing is efficient and compared the contribution of pre-processing in constructing a gas-poor galaxy population among our YZiCS clusters. Our discussions in this section are summarized as follows.

In summary, we have demonstrated that $M_{\rm h, before}$ (the average of host halo masses before the cluster infall) is a better indicator than $M_{\rm h,infall}$ (the host halo mass at the time of the cluster entry) in estimating the degree of pre-processing. Pre-processing is found to be more effective on a {\em smaller satellite} galaxy in a {\em more massive} group, which appears to be consistent with what is expected from the ram pressure stripping process. Finally, pre-processing is more significant in more massive clusters primarily because massive clusters are assembled through more massive groups in the hierarchical paradigm.

%\begin{enumerate}[label={(\roman*)}]
%    \item When using the host halo mass at the time of the infall ($M_{\rm h,infall}$) as an indicator of the environmental effect prior to the cluster infall, it often misestimates the integrated history. Instead, we use the average of all host halo masses before the infall ($M_{\rm h, before}$) that provide a better indication of the previous environments to demonstrate the effect of group environments on producing gas-poor galaxies.
    
%    \item Pre-processing is more efficient for lower mass satellite galaxies in more massive halos, especially when the mass ratio (the stellar mass of a satellite divided by the mass of its host group halo) is small. This is consistent with what is expected from ram pressure stripping. We present more evidence for ram pressure stripping in group-size halos in Section \ref{s6.1}.
    
%    \item
%    Massive clusters usually grow in dense environments crowded with rich and massive groups. As a result, they are more likely to accrete galaxies that have undergone group environmental effects prior to the cluster infall. Pre-processing plays an important role in constructing the gas-poor galaxy population in massive clusters.
%\end{enumerate}

\section{Gas depletion inside cluster halo} \label{s5}

In the previous section, we discovered that a non-negligible fraction of galaxies ($\sim30\%$) arrive at the cluster as gas-poor galaxies. 
%Yet, $\sim70\%$ of galaxies obtain accreted clusters with gas, and the fraction is seemingly higher in less massive clusters. 
This means that the majority of galaxies are still gas rich at the cluster entry. In this section, we focus on the processes affecting the gas of these galaxies after arriving at clusters. In particular, our main interest is the effect of ram pressure, which is considered as a leading mechanism that induces a severe environmental effect on satellite galaxies falling into massive halos (e.g., \citealt{Koopmann_2004}; \citealt{Tonnesen_2007}; \citealt{Tecce_2010}; \citealt{Kimm_2011}; \citealt{Vollmer_2012}).

\subsection{How violent is the depletion inside clusters?} \label{s5.1}

\begin{figure*}[th]
    \centering
    \includegraphics[width=0.9\textwidth]{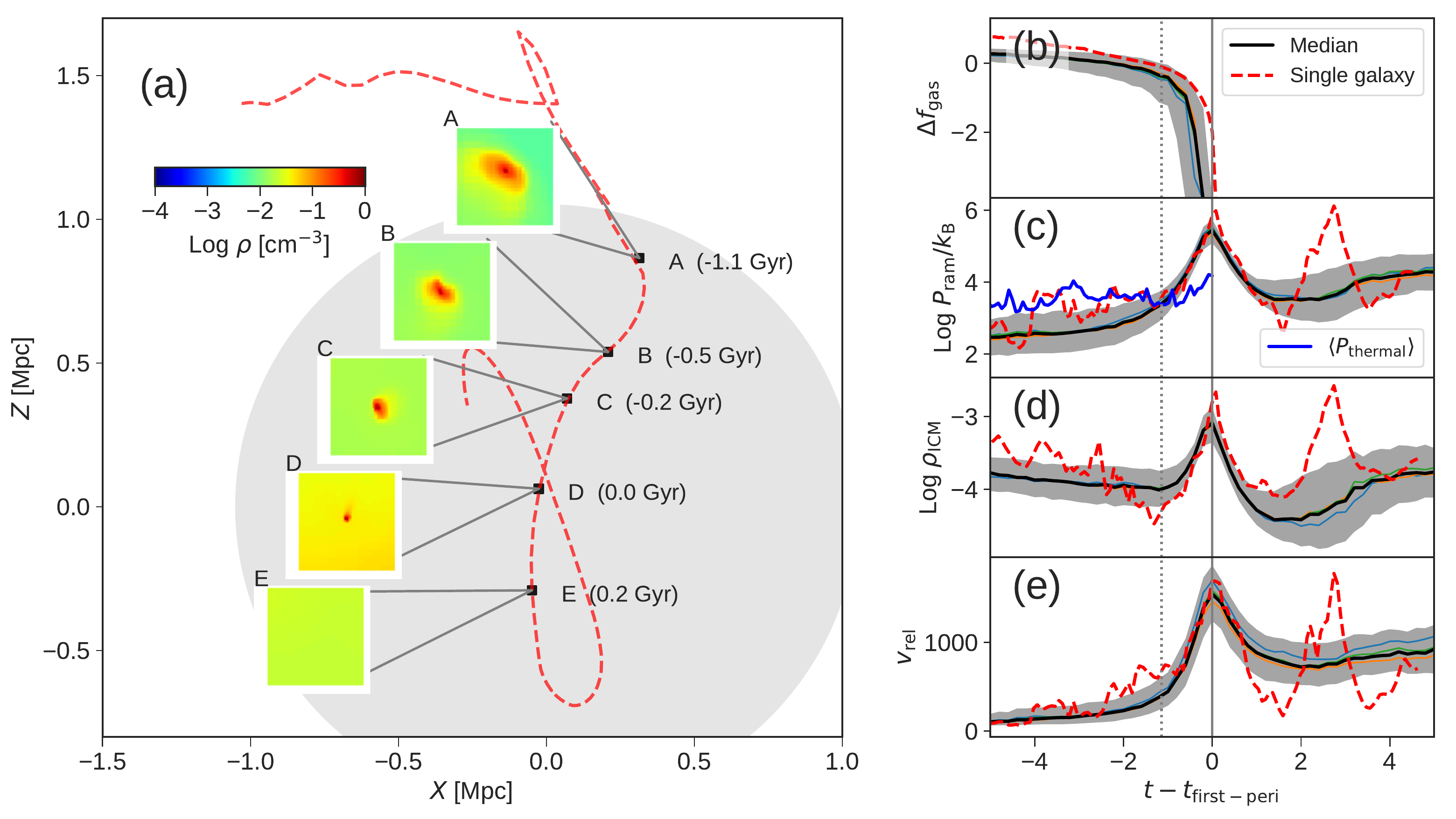}
    \caption{
    Evolution of a gas-rich galaxy falling into a cluster. (a) The red dashed line is the infalling orbit and the panels labeled from A to E show the transformation in the gas morphology. The gray circle shows the cluster region ($< 1.5\, R_{\rm 200}$), and numbers inside brackets are the time before (negative) or after (positive) the first cluster pericentric pass. This galaxy underwent severe gas stripping, losing the majority of its gas during the first approach to the cluster center (see also Table \ref{tab:t1}). The panels on the right show the time evolution of the (b) gas content (c) magnitude of the ram pressure [$\rm cm^{-3}\,\rm K$], (d) ambient medium density [$\rm cm^{-3}$], and (e) velocity of a galaxy with respect to its ambient medium [$\kms$]. The x-axis is synchronized to 0 at the first pericentric passage moment of the galaxies. The red dashed lines show the evolution of the example galaxy shown in panel (a), solid black lines are the median of all the galaxies, and gray shades show the 30th and 70th percentiles. The evolution of gas content of galaxies well follows the change in the ram pressure and the strong ram pressure experienced by the example galaxy is common among our YZiCS galaxies.
    }
    \label{fig:f10}
\end{figure*}

\begin{table*}[!]
\centering
 \begin{threeparttable}
 \caption{Time evolution of gas properties of a sample galaxy displayed in Figure~\ref{fig:f10}.}
 \label{tab:t1}
 \begin{tabular}{clcccccc}
  \hline
    & & t - $t_{\rm first-peri}$ [Gyr] & $M_{\rm gas}$ [$\rm M_{\rm \odot}$] & $M_{\rm gas}$ / $M_{\rm gas,infall}$ & $P_{\rm ram}/k_{\rm B}$ [$\rm cm^{-3}\,\rm K$] & $\langle P_{\rm thermal}/k_{\rm B} \rangle$ [$\rm cm^{-3}\,\rm K$]\\ \hline
  A & Cluster infall & -1.06 & 3.9 $\times$ 10$^{9}$ & 1.0 & 5.5 $\times$ 10$^{3}$ & 2.3 $\times$ 10$^{3}$ \\ \hline
  B & & -0.49 & 1.5 $\times$ 10$^{9}$ & 0.375 & 2.6 $\times$ 10$^{4}$ & 7.8 $\times$ 10$^{3}$\\ \hline
  C & & -0.24 & 5.6 $\times$ 10$^{8}$ & 0.141 & 7.2 $\times$ 10$^{4}$ & 6.9 $\times$ 10$^{3}$\\ \hline
  D & The first pericenter & 0.0 & 5.9 $\times$ 10$^{7}$ & 0.015 & 7.2 $\times$ 10$^{5}$ & 1.5 $\times$ 10$^{4}$ \\ \hline
  E & & +0.24 & 0.0 & 0.0 & 1.7 $\times$ 10$^{5}$ & - \\ 
  \hline
 \end{tabular}

 \begin{tablenotes}
  \small
   \item Note. The first column corresponds to each snapshot marked on the panel (a) of Figure~\ref{fig:f10}. The columns from left to right are the (1) time before and after the pericentric pass, (2) cold gas mass, (3) cold gas mass scaled with the gas mass at the infall, (4) magnitude of the ram pressure, and (4) mean thermal pressure of the cold gas cells that belong to the galaxy.
 \end{tablenotes} 
 \end{threeparttable}
\end{table*}

As discussed in Section \ref{s3}, there is a strong environmental effect acting on our YZiCS galaxies. Figure~\ref{fig:f10} shows one example of a gas-rich galaxy falling into a cluster. The small figures in panel (a) are gas density maps at given snapshots during the first infall (labeled A, B, C, D, and E). We can see this galaxy loses a significant amount of gas during its first infall, starting from the outskirt, and none of its cold gas is detected at snapshot E. By measuring the cold gas mass at each moment (see columns (2) and (3) of Table \ref{tab:t1}), we find that the galaxy loses $99.5\%$ of its initial cold gas during its first approach to the center.

We measure the ram pressure intensity that each galaxy goes through following the \citet{Gunn&Gott_1972} description:
\begin{equation}
P_{\rm ram} = \rho_{\rm ICM} v_{\rm rel}^{2}\,,
\end{equation}
where $\rho_{\rm ICM}$ is the mass-weighted average gas density of the ambient medium cells defined in Section \ref{s2.3.1}, and $v_{\rm rel}$ is the velocity of a galactic center with respect to the mean motion of the ambient medium. With our definition of the ambient medium, we do not take account of the possible ram pressure that could be caused by other galaxies as the galaxy passed through the neighboring gas.

When galaxies are at their first pericenter, the ram pressure acting on the galaxies is at its maximum strength. The red dashed lines in panels (b), (c), (d) and (e) follow the time evolution of gas content, ram pressure, ambient hot gas density, and velocity of a galaxy with respect to its surrounding hot gas. 

The gas content parameter of a galaxy, $\Delta f_{\rm gas}$, is defined as
\begin{equation} \label{eq:deficiency}
{\Delta f_{\rm gas} = \log \, f_{\rm gas} - {\langle\log \, f_{\rm gas}\rangle}_{\rm *} }\, ,
\end{equation}
where $f_{\rm gas} \equiv M_{\rm gas}/M_{\rm *}$ and $\langle\log \, f_{\rm gas}\rangle_{\rm *}$ is the mean gas fraction of the gas-rich galaxies at the stellar mass given on the $\log \, M_{\rm *} - \log \, f_{\rm gas}$ plane (the black solid line on Figure~\ref{fig:f3}). As a galaxy undergoes the gas depletion process, $\Delta f_{\rm gas}$ decreases and moves downward in the $\log \, M_{\rm *} - \log \, f_{\rm gas}$ plane. When $\Delta f_{\rm gas}$ is lower than -1.07, the 2$\sigma$ clipping result, the galaxy is defined as a gas-poor galaxy following the definition introduced in Section \ref{s2.3.2}. We hereby define $\Delta f_{\rm gas, infall}$ as the gas content parameter of a galaxy at the time it arrived at 1.5$\,R_{\rm 200}$ of the cluster for the first time.

The x-axis in this figure is the time synchronized at their first pericenteric passage moment at 0. It is evident that the measurements in panels (c), (d), and (e) peak at the pericenter. At the peak, the ram pressure reaches $P_{\rm ram} / k_{\rm B} \sim 10^6\, \rm cm^{-3} K$ which is two orders of magnitude larger than what it was at the time the galaxy first cross 1.5$\,R_{\rm 200}$ of the cluster halo (the vertical dotted line).

In order to estimate the efficiency of the ram pressure stripping, we compare the mass-weighted average of thermal pressures of all the cold gas cells belong to a galaxy ($\langle P_{\rm thermal} \rangle$) with the magnitude of the ram pressure. Assuming that galactic gas cells are in a hydrostatic state, their thermal pressure balances with the gravity at the position. Thus when the ram pressure acting on a galaxy is stronger than $\langle P_{\rm thermal} \rangle$, the galactic gas is pushed out from the gravitational potential (see \citealt{Mori_2000}; \citealt{Marcolini_2003}; \citealt{Roediger_2005}). The cold gas cells of the sample galaxy have their average thermal pressure of $\langle P_{\rm thermal}/k_{\rm B} \rangle\sim10^{3-4} \rm cm^{-3}\,K$ at all times (see the blue solid line in panel (b)), so that gas stripping is expected to occur ($\langle P_{\rm thermal} \rangle < P_{\rm ram}$) once the galaxy falls inside roughly $1.5\,R_{\rm 200}$ of the cluster halo. Considering that the thermal pressure is highest at the central region of a galaxy and decreases going outward, some gas stripping can take place at the outskirts of the galaxy even earlier. We will show in Section \ref{s6.1} that the amount of stripped gas increases with the magnitude of the ram pressure, starting from $P_{\rm ram} / k_{\rm B} \sim 10^{4}$ $\rm cm^{-3}\,\rm K$.

\begin{figure}
    \centering
    \includegraphics[width=0.9\columnwidth]{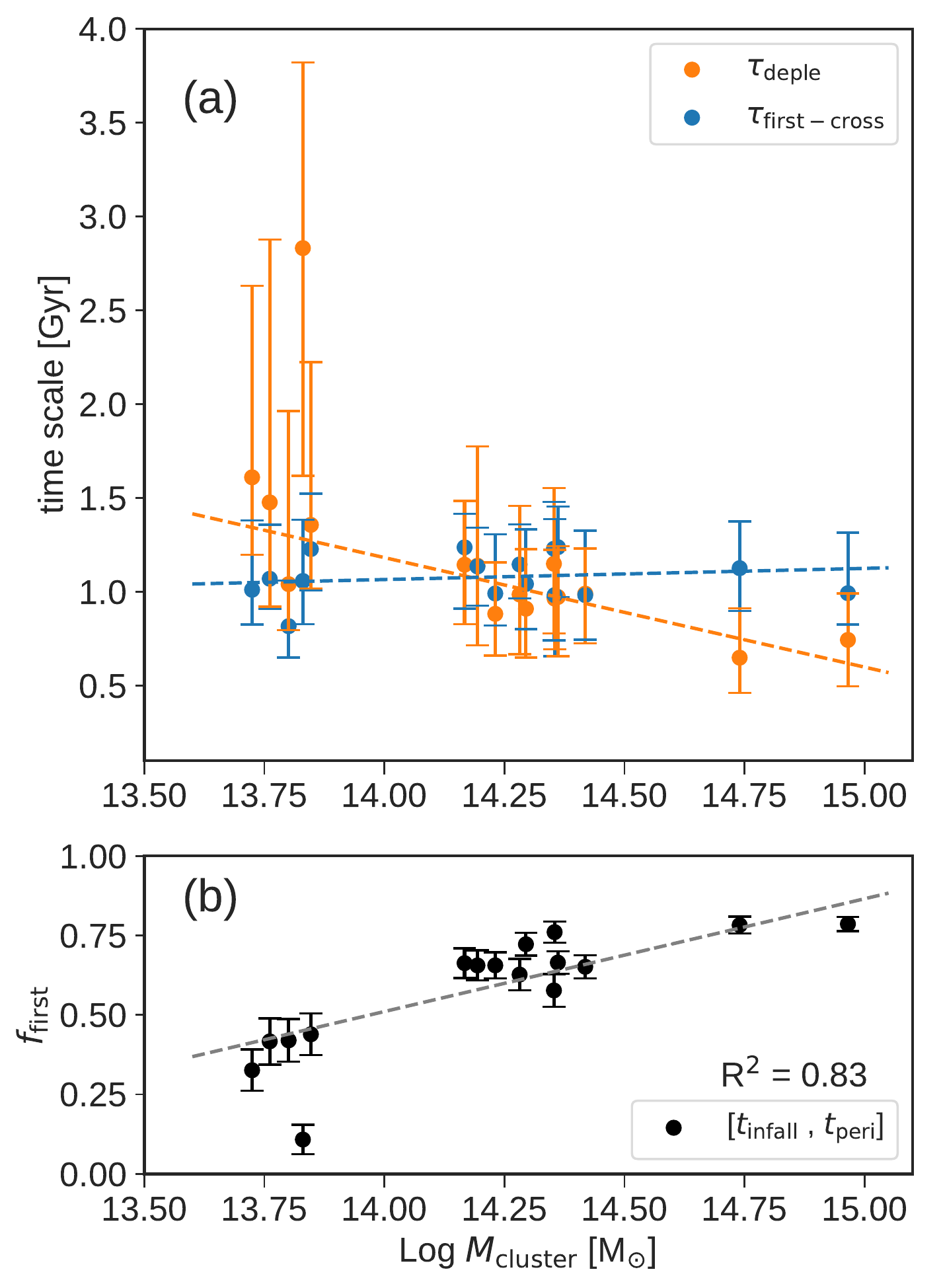}
    \caption{
    (a) $\tau_{\rm deple}$ (orange) and $\tau_{\rm first-cross}$ (blue) as a function of cluster halo mass. The circular symbols and error bars represent the medians and the 30th and 70th percentiles. $\tau_{\rm deple}$ decreases with increasing cluster mass, while $\tau_{\rm first-cross}$ remains constant. $\tau_{\rm deple}$ is generally shorter than $\tau_{\rm first-cross}$ in massive clusters ($\gsim 10^{14.5} \rm M_{\rm odot}$) and the opposite in low mass clusters ($\lsim 10^{14} \rm M_{\rm odot}$). (b) Fraction of galaxies that become gas poor during their first infall, i.e., $\tau_{\rm deple} < \tau_{\rm first-cross}$, as a function of cluster halo mass (only the galaxies that reach their first pericentric pass by redshift 0 are used here). Error bars are the standard error of the mean. The least square linear fit to the data is shown as dashed line, and its associated Pearson coefficient is given in Panel (b).
    }
    \label{fig:f11}
\end{figure}

Note that the time evolution of gas content parameter in panel (b) follows the change in the ram pressure. If other environmental processes such as galaxy interactions played an important role in gas stripping, the time evolution of gas content parameter would appear stochastic instead of smooth. This strongly suggests that {\em ram pressure is the dominant source of gas stripping inside clusters}.

%As seen in the example galaxy, the strong ram pressure effects seem to be common among our galaxies (see the gray shade in panel (c) indicating the 30th and 70th percentiles of the ram pressure intensity distribution). This implies that the severe gas loss during the first infall is a common process. 

We examine the fraction of galaxies that become gas poor {\em during} the first infall by measuring the time for galaxies to become gas poor ($\tau_{\rm deple} \equiv t_{\rm deple} - t_{1.5\, R_{\rm 200}}$, termed depletion timescale) and compare it with a galaxy's first-crossing timescale ($\tau_{\rm first-cross}\equiv t_{\rm first-peri} - t_{1.5\, R_{\rm 200}}$)\footnote{For ``crossing time'', we take the time for a galaxy to reach from 1.5$\, R_{\rm 200}$ to its first pericenter because ram pressure stripping seems to be effective starting from 1.5$\, R_{\rm 200}$ in our simulation, as discussed above in this section and panel (a) of Figure~\ref{fig:f10}}. In panel (a) of Figure~\ref{fig:f11}, the orange symbols show the median depletion timescale of galaxies of each cluster and the error bars are the 30th and 70th percentiles. The satellite galaxies in more massive clusters lose their cold gas faster (see the linear fit; orange dashed line), i.e., the environmental effect is stronger in more massive clusters. On the contrary, the first-crossing timescale (blue symbols and error bars) is nearly constant over the range of cluster mass when stacking galaxies with various infall times, which is consistent with the simple expectation for the dynamical timescale of virialized systems ($\tau_{\rm dyn} \propto 1/\sqrt{G \rho}$, where $\rho$ is the overdensity of a dark matter halo which is 200 times greater than the critical density of the universe). %Therefore, we expect that galaxies falling into more massive clusters are more likely to become gas poor during their first approach to the center.
Panel (b) shows the fraction of galaxies that become gas poor during the first infall ($\tau_{\rm deple}/\tau_{\rm first-cross}$ < 1) in each cluster.  As expected from the panel (a), the fraction increases with increasing cluster mass. Nevertheless, a notable fraction of galaxies ($\sim35\%$) has their gas depleted by the first pericentric pass even in the low-mass clusters ($< 10^{14}\, \rm M_{\rm \odot}$), i.e., the severe gas loss found in the previous example (Figure~\ref{fig:f10} and Table \ref{tab:t1}) is not exclusive to massive cluster environments only.

%In Section \ref{s3}, we defined ``the first pericentric pass'' of a galaxy as the moment when the moving direction of a galaxy becomes perpendicular to the position vector pointing the cluster center for the first time. By definition, $\tau_{\rm first-cross}$ is measured until a galaxy reaches its first pericenter, i.e., when the ram pressure acting on the galaxy is strongest. Considering that there are galaxies that deplete their gas right after their pericentric pass, it is worthwhile to examine the changes in the gas-poor galaxy fraction by allowing extra time in the measurement window. The open circles show the fraction of gas-poor galaxies between the cluster infall moment and the additional 1$\Gyr$ after the pericenter. The overall trend is identical to the filled circles and thus robust but shows a $\sim0.2\rm dex$ increase in the fraction. 
%Therefore, this cluster mass trend is valid without concerning the exact definition of the end of the first infall.

\subsection{What determines the gas depletion timescale in clusters?} \label{s5.2}

%As the first-crossing timescales of infalling galaxies are almost identical in all cases, it is the depletion timescale of a galaxy that determines whether or not a galaxy becomes gas poor during the first infall.
In this section, our aim is to answer %how long galaxies take to become gas poor in cluster environments and
in what conditions gas depletion is efficient (see also \citealt{McCarthy_2008}). We base our analysis on the following three parameters representing the physical properties of infalling galaxies; gas content, stellar mass and orbital shape.

%The gas content parameter of a galaxy, $\Delta f_{\rm gas}$, is introduced to be  
%\begin{equation} \label{eq:deficiency}
%{\Delta f_{\rm gas} = \log f_{\rm gas} - {\langle\log f_{\rm gas}\rangle}_{\rm *} }\, ,
%\end{equation}
%where $f_{\rm gas} \equiv M_{\rm gas}/M_{\rm *}$ and $\langle\log f_{\rm gas}\rangle_{\rm *}$ is the mean gas fraction of the gas-rich galaxies at the galaxy mass given on the $\log M_{\rm *} - \log f_{\rm gas}$ plane (the black solid line on Figure~\ref{fig:f3}). As a galaxy undergoes the gas depletion process, $\Delta f_{\rm gas}$ decreases and moves downward in the $\log M_{\rm *} - \log f_{\rm gas}$ plane. When $\Delta f_{\rm gas}$ is lower than -1.07, the 2$\sigma$ clipping result, the galaxy is defined as a gas-poor galaxy following the definition introduced in Section \ref{s2.3.2}. We hereby define $\Delta f_{\rm gas, infall}$ as the gas content parameter of a galaxy at the time it arrived at 1.5$\,R_{\rm 200}$ of the cluster for the first time.

\begin{figure}
    \centering
    \includegraphics[width=0.9\columnwidth]{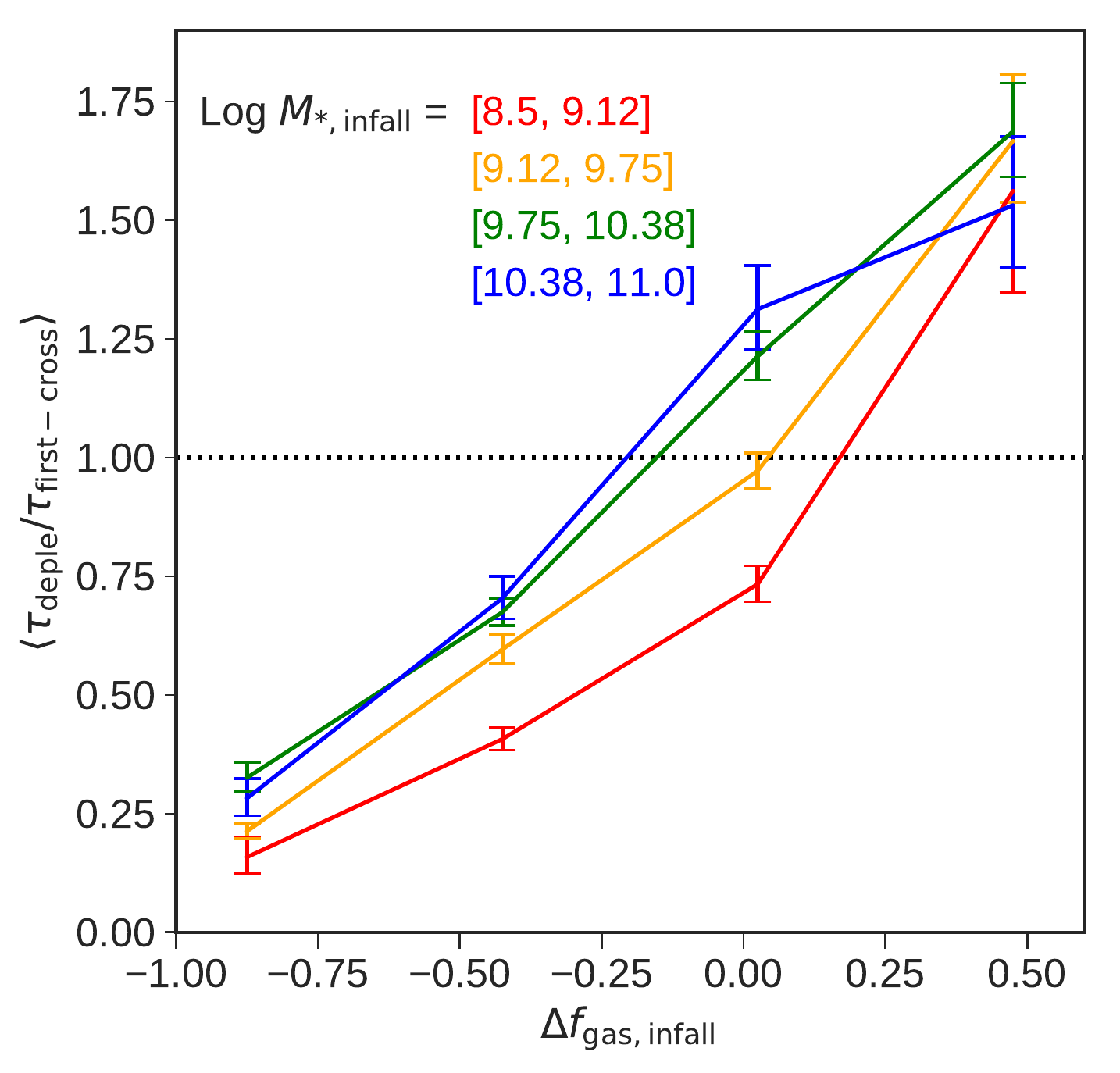}
    \caption{
    $\tau_{\rm deple} / \tau_{\rm first-cross}$ versus $\Delta f_{\rm gas, infall}$. Each line represents the galaxies in different $M_{\rm *, infall}$ ranges at the time of the infall. At all stellar mass ranges, galaxies take a longer time to be gas poor with increasing $\Delta f_{\rm gas, infall}$. At fixed $\Delta f_{\rm gas, infall}$, lower mass galaxies deplete their gas faster, but the dependence on the stellar mass is weaker than on $\Delta f_{\rm gas, infall}$. The black horizontal line marks where $\tau_{\rm deple}/\tau_{\rm first-cross} = 1$, i.e., where the gas depletion timescale equals the first-crossing timescale. The galaxies under this line become gas poor during their first infall.
    }  
    \label{fig:f12}
\end{figure}

\begin{figure}
    \centering
    \includegraphics[width=0.9\columnwidth]{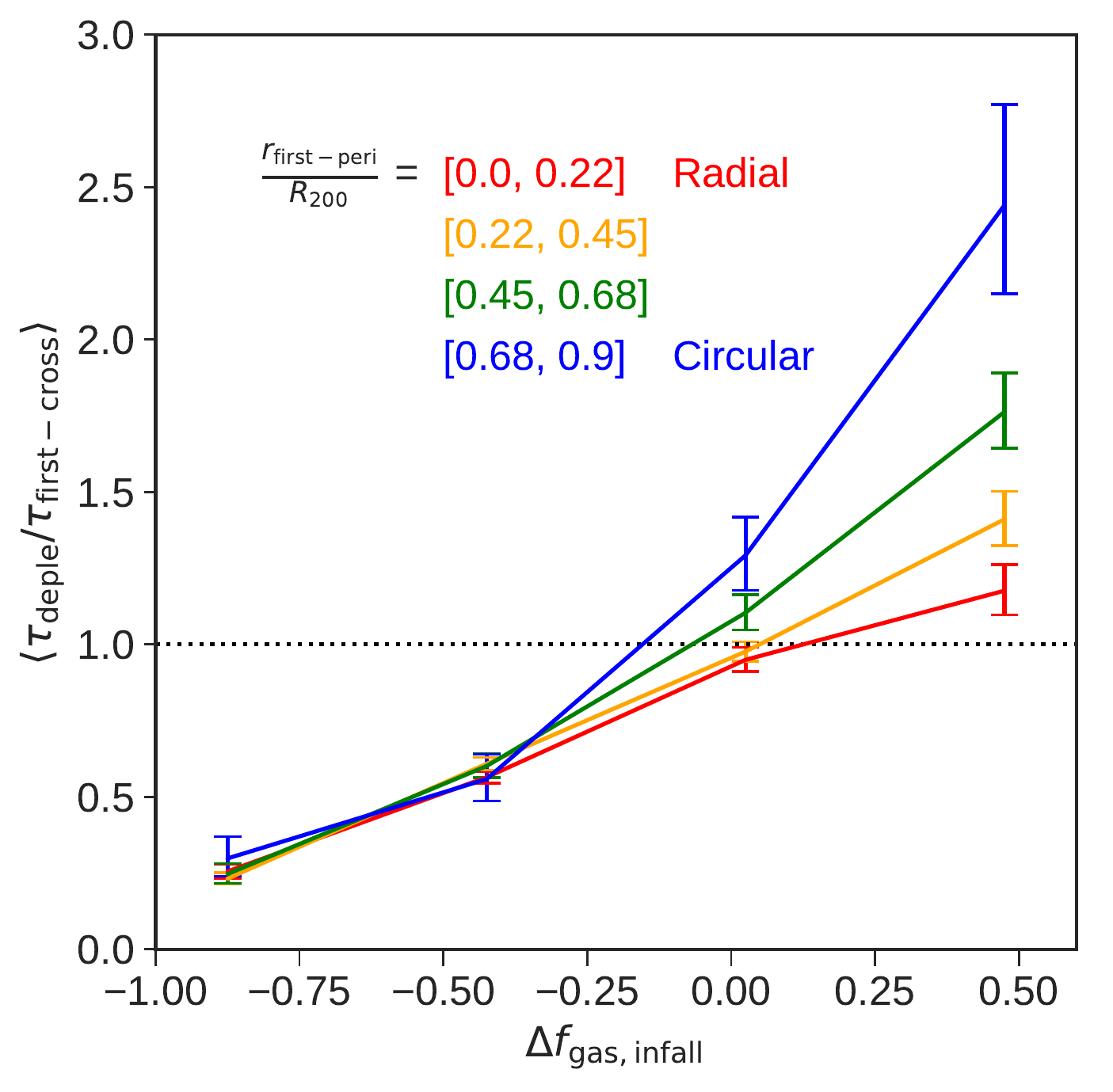}
    \caption{
    The same format as for Figure~\ref{fig:f12}, but each color corresponds to a different shape of the orbit ($r_{\rm peri} / R_{\rm 200}$). The galaxies with low $\Delta f_{\rm gas, infall}$ become gas poor quickly after they arrive at clusters almost independent of the orbital shape. When $\Delta f_{\rm gas, infall}$ is high, galaxies take a longer time to be gas poor and the gas depletion timescale strongly depends on the shape of the infalling orbit: galaxies with radial orbit lose their gas faster.
    }
    \label{fig:f13}
\end{figure}

Figure~\ref{fig:f12} shows $\tau_{\rm deple} / \tau_{\rm first-cross}$ depending on $\Delta f_{\rm gas, infall}$, and each line represents the galaxies in different $M_{\rm *, infall}$ ranges. Note that we present the depletion timescales averaged in log scale as their distribution better follows a lognormal distribution, rather than a Gaussian. The most prominent trend shown here is that it takes longer for the galaxies with higher values of $\Delta f_{\rm gas, infall}$ (i.e., gas-rich galaxies) to become gas poor. 
%If a galaxy already has a low $\Delta f_{\rm gas, infall}$, it is easier for it to become gas-poor compared to galaxies with higher $\Delta f_{\rm gas, infall}$.

At fixed $\Delta f_{\rm gas, infall}$, the gas depletion is slightly faster when the stellar mass of galaxies is smaller. This trend agrees with the findings on pre-processing in Section \ref{s4}; that is, galaxies with lower stellar masses are more likely to become gas poor within group halos. Note that our definition of gas-poor galaxies is applied at a fixed stellar mass in order to consider the stellar mass dependence of the gas fraction. Therefore, the discrepancy in the depletion timescale between low and high mass galaxies can be considered as a distinction in the efficiency of the gas depletion process in clusters. 

In Figure~\ref{fig:f13}, we compare the effect of $\Delta f_{\rm gas, infall}$ and the orbital shape on the gas depletion timescale. As a proxy of the orbital shape, we use the first pericentric distance of galaxies ($r_{\rm first-peri}$/$R_{\rm 200}$), where more radial orbits approach closer to the cluster center. The dependence of depletion timescale on $\Delta f_{\rm gas, infall}$ shown in Figure~\ref{fig:f12} is also visible here, but the slope of the trend varies with the shape of the orbit. Considering galaxies with high $\Delta f_{\rm gas, infall}$, galaxies in radial orbits (red line) deplete their gas much faster as they approach closer to the cluster center, while those with more circular orbits (blue line) need a longer time to be gas poor. When it comes to the relatively low $\Delta f_{\rm gas, infall}$ galaxies the dependence of depletion timescale on the orbital shape is weakened and therefore majority of the galaxies, regardless of their orbit, become gas deficient during their first infall ($\tau_{\rm deple} / \tau_{\rm first-cross}$ < 1). Among our sample, the most common orbit is $r_{\rm first-peri}/R_{\rm 200}\sim 0.3$ regardless of the cluster mass, i.e., majority of the galaxies take a radial orbit at their first approach to the center (see also \citealt{Ghigna_1998}; \citealt{Tormen_2004}; \citealt{Benson_2005}).

\subsection{What determines the initial gas content?}
\begin{figure*}
    \centering
    \includegraphics[width=0.9\textwidth]{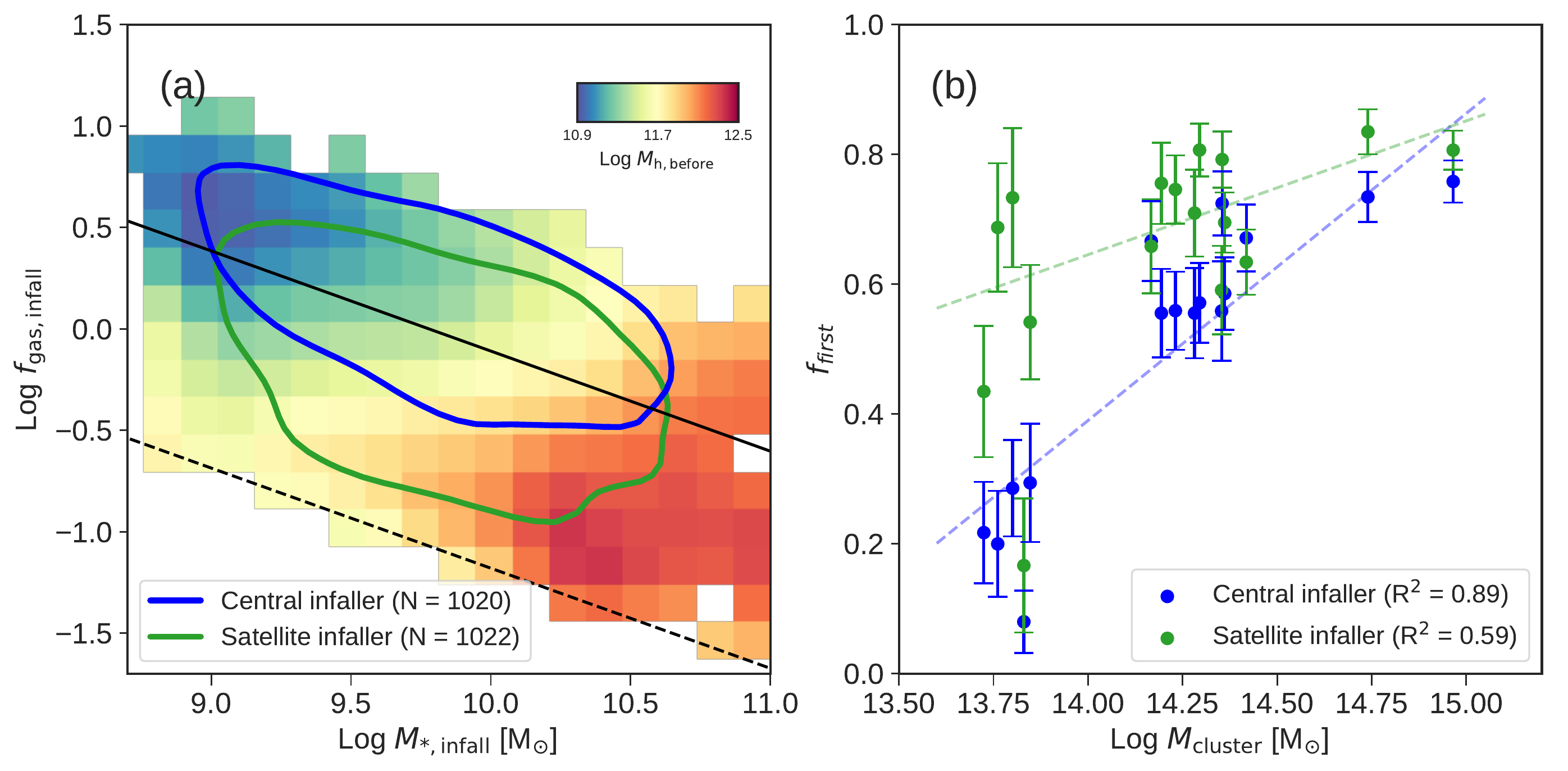}
    \caption{
    (a) The 1$\sigma$ distribution of central (blue) and satellite (green) galaxies on the $M_{\rm *, infall}$ - $f_{\rm gas,infall}$ plane. The black solid line is the gas-rich galaxy sequence at redshift 0 presented in Figure~\ref{fig:f3} and the dashed line is the demarcation of the gas-poor galaxies. At a fixed stellar mass, the satellite infallers have a lower gas fraction compared to centrals at the time they first arrive at clusters. The background colors indicate the average $M_{\rm h, before}$ at each bin. The galaxies with lower gas fraction galaxies tend to have higher $M_{\rm h, before}$. It was the group environmental effects before the cluster entry that caused the variation in the gas fraction at the time of the infall of our galaxies. (b) The fraction of galaxies that become gas deficient during their first infall as a function of the cluster mass (only the galaxies that reach their first pericentric pass by redshift 0 are used). Error bars are measured as the standard error of the mean of a binomial distribution. The galaxies are separated into groups of the central (blue) and satellite infallers (green). At all clusters, the satellite infallers are more likely to become gas poor during their first infall.
    }
    \label{fig:f14}
\end{figure*}

We have shown in the previous section that $\Delta f_{\rm gas, infall}$ is the primary parameter that determines the gas depletion timescale. Now, we examine the origin of its variation and its consequence.

We divided cluster-infalling galaxies into central and satellite infallers following the definitions in Section \ref{s4}, i.e., galaxies that have always been central galaxies before reaching clusters and galaxies that have been a satellite. In Figure~\ref{fig:f14}, we found central infallers generally (blue solid contour) have larger gas content ($f_{\rm gas,infall}$) than satellite infallers (green solid contour) at fixed stellar mass at the time they arrive at clusters (see also \citealt{Catinella_2013}). As introduced in Figure~\ref{fig:f3}, the black solid and dashed lines represent the gas-rich galaxy sequence and the demarcation of the gas-poor galaxies, respectively. 
%One might wonder on the origin of the gas poor nature of the satellite infallers. We here show their past values of $f_{\rm gas}$ when they became a satellite of a large (group) halo for the first time (green dashed contour). They were not different from normal central galaxies in our sample in terms of gas fraction, except that they were richer in gas simply because the typical gas fraction of galaxies was higher in the past than today.
%By investigating $f_{\rm gas}$ of satellite infallers at the time they first became satellites of larger halos (green dashed contour), we find satellite infallers were gas-rich at that time. 
The galaxies with lower values of $f_{\rm gas,infall}$ have higher $M_{\rm h, before}$ at fixed stellar mass (see the background colors in each bin).
We are thus confident that it was the environmental effect in previous group halos that caused the low values of $f_{\rm gas,infall}$ of our satellite infallers.
%in group halos that induces the distinction in $f_{\rm gas}$ among our galaxies. In other words, even though the satellite galaxies affected by group-size halos do not completely run out of gas before the cluster entry, i.e., are completely pre-processed, their gas fraction is comparably lower compared to the central galaxies.

Combining our findings that (i) galaxies with a lower gas content at the time of their infall lose their gas faster and (ii) the gas content of satellite galaxies are lower than those of central galaxies, we understand the clear separation between the central and satellite infallers in terms of the fraction of galaxies that become gas poor during the first infall (see panel (b) of Figure~\ref{fig:f14}). Generally, the satellite infallers (green) show a larger fraction of galaxies becoming gas-poor galaxies before their first pericentric pass compared with the central infallers (blue). Also, note that the slope of the linear fit is different between the two groups. In massive clusters ($> 10^{14.5}\, \rm M_{\rm \odot}$), both satellite and central infallers go through the extreme environmental effect, capable of depleting their cold gas during their first approach to the center. When it comes to low mass clusters ($< 10^{14}\, \rm M_{\rm \odot}$), the environmental effect in these clusters is still sufficient to deplete the satellite infallers, but not strong enough to deplete the central infallers during their first infall. Therefore, we can say that cluster mass dependence on the efficiency of environmental effect shown in panel (b) of Figure~\ref{fig:f11} is mainly caused by the central infallers.

\subsection{Summary on cluster process}

In this section, we discuss the violent nature of the gas depletion of galaxies falling into clusters, in particular during the first approach to the center. We also examine the effect of the physical properties of infalling galaxies, e.g., stellar mass, gas fraction, and orbital shape, on the gas depletion timescale. Generally, our galaxies significantly lose their gas (starting from the outskirts) during the first approach to the cluster center. The following is a summary of the results.

\begin{enumerate}[label={(\roman*)}]
    %\item The environmental effect acting on a galaxy, e.g., ram pressure intensity, peaks at the pericenter. Generally, our galaxies significantly lose their gas (starting from the outskirts) and display disturbed gas morphology during the first approach to the cluster center. We find that a larger fraction of galaxies become gas poor during their first infall in a more massive cluster.

    \item Gas depletion timescale strongly depends on the gas content of galaxies at the time of cluster entry. The larger the gas content, the longer it takes to be ``gas poor''. 
    %Galaxies that arrive at clusters with a lower gas fraction lose their remaining gas faster and become gas poor during their first infall more easily. 
    %Lower mass galaxies take shorter time to become gas poor, but gas fraction has a greater impact on the gas depletion timescale.
    Galaxy mass also affects gas depletion timescale in the sense that lower mass galaxies take a shorter time to become gas poor, but its impact is smaller than that of gas content.
    %is more sensitive to the gas fraction than to the stellar mass. 
    %the effect of the stellar mass on the depletion timescale is weaker than the effect of gas fraction at the time of the infall.
    
    \item The orbital shape affects the gas depletion timescale of a galaxy. The galaxies falling with a more radial orbit lose their gas faster. However, the effect of orbital shape becomes weaker when the infalling galaxy is low in gas content. 
    %In this case, a galaxy becomes a gas-poor galaxy during the first approach to the center regardless of the orbit.
    
    \item By separating the central and satellite infallers, we find a distinction between the two populations in terms of the gas content at the time they arrive at the clusters: central infallers on the average have a higher gas content. We claim that the variation in gas content originates from the environmental effect in group-size halos prior to the cluster entry.
    
    \item Satellite infallers (with a lower gas content) take a shorter time to become gas poor almost regardless of the cluster mass. On the contrary, central infallers (with a higher gas content) show a strong dependence on cluster mass.
    %because it takes a stronger environmental effect to remove such a large amount of gas.
    %Satellite infallers are more likely to become gas poor during their first infall regardless of the cluster mass, while central infallers have a stronger dependence on cluster mass. This is mainly due to the difference in gas content between satellite and central galaxies.
\end{enumerate}

\section{Discussion} \label{s6}

\subsection{Ram pressure stripping in YZiCS galaxies} \label{s6.1}

\begin{figure*}
    \centering
    \includegraphics[width=0.9\textwidth]{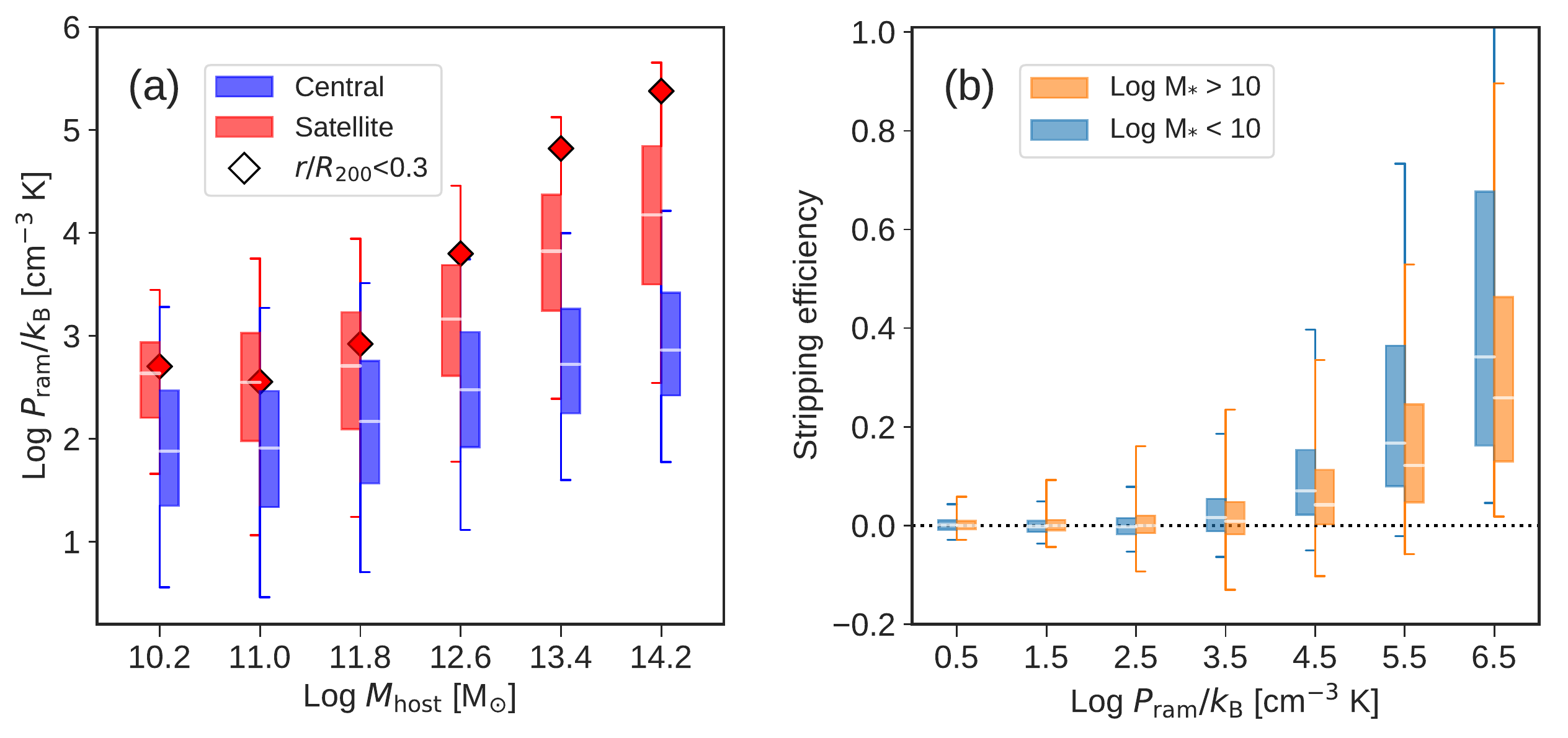}
    \caption{
    (a) The ram pressure intensity acting on galaxies versus their host halo mass. The bottom and top of each box are the first and third quartiles, and the whiskers show the 5th and 95th percentiles. The red and blue colors represent the satellite and central galaxies, respectively. The red diamond symbol shows the median ram pressure intensity of satellite galaxies in the inner region ($< 0.3\,R_{\rm 200}$) of their host halos. Satellite galaxies always experience stronger ram pressure (especially satellites in the inner regions) compared with centrals and the ram pressure acting on satellites become stronger with increasing host mass. (b) Gas stripping efficiency of galaxies, i.e., the amount of gas stripped out between two snapshots divided by the initial gas mass (see equation \ref{eq:def_gas_loss}), versus the ram pressure intensity. The blue and orange colors indicate the less massive ($< 10^{10}\,\rm M_{\rm \odot}$) and massive ($> 10^{10}\,\rm M_{\rm \odot}$) galaxies, respectively. With increasing ram pressure intensity, galaxies strip their gas more efficiently between snapshots. In particular, less massive galaxies with lower gravitational restoring force lose their gas more easily than massive galaxies at a given ram pressure intensity.
    }  
    \label{fig:f15}
\end{figure*}

\begin{figure*}
    \centering
    \includegraphics[width=0.9\textwidth]{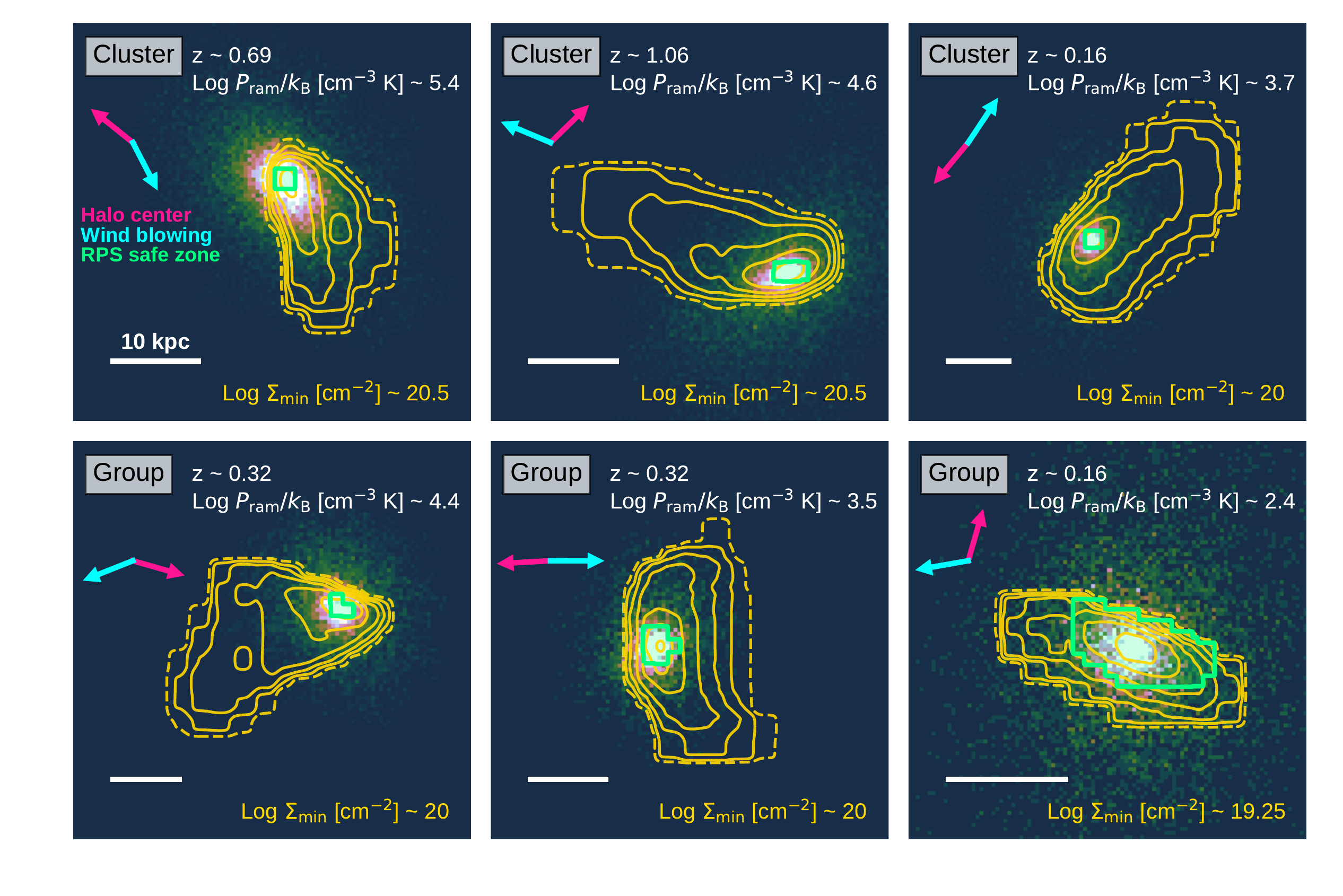}
    \caption{
    Distribution of star particles (background color) and gas column density (yellow contour) on the projected plane for six example galaxies in a different level of ram pressure at different redshift. Each contour line is separated by 0.4 dex in log density and the dashed line corresponds to the minimum level presented in the bottom right of each panel. Galaxies in the upper panels are selected among the galaxies inside clusters and bottom panels show the group galaxies outside clusters. The white bar indicates the 10$\kpc$ length scale. The pink and blue arrows indicate the direction to the host halo center and the wind-blowing direction, i.e., the opposite direction with a galaxy's motion relative to the ambient medium, respectively. The diffuse gas components extend toward the wind-blowing direction, while the stellar components remain unaffected. The green colored region shows the distribution of ``ram pressure stripping safe'' cells, i.e., gas cells with their thermal pressure higher than the ram pressure acting on the galaxy ($P_{\rm thermal} > P_{\rm ram}$). These cells reside at the central region of the galaxies.
    }  
    \label{fig:f16}
\end{figure*}

In Section \ref{s5.1}, we demonstrated that the pattern of gas stripping is well matched by the characteristics of ram pressure stripping. We here discuss it in further detail.

%Although there are variety of possible interactions in dense environments that may cause gas loss from galaxies, e.g., tidal interaction, comparing the relative dominance between them is beyond the scope of this paper. Instead, we investigate how well the trends expected from ram pressure stripping process appears among our sample and present evidence that ram pressure stripping is likely to play an important role in depleting cold gas both in clusters and group-size halos. 

Panel (a) of Figure~\ref{fig:f15} demonstrates how the magnitude of the ram pressure ($P_{\rm ram} = \rho_{\rm ICM} v_{\rm rel}^{2}$) acting on galaxies varies with host halo mass. The bottom and top of each box are the first and third quartiles, and the whiskers show the 5th and 95th percentiles.
Since the intensity of the ram pressure can vary significantly depending on the radial distance from the halo center, we also show the ram pressure intensity measured only with the satellite galaxies residing in the central region ($< 0.3\,R_{\rm 200}$) of their host halos with diamond symbols. Satellite galaxies (red) always experience stronger ram pressure compared to centrals. In particular, in massive groups ($> 10^{12}\,\rm M_{\odot}$), the ram pressure acting on the satellites in the central regions (red diamonds) is about two orders of magnitude stronger than that acting on in the centrals in the same halos. As the host halo mass increases, satellite galaxies undergo stronger ram pressure because of the increased intra-group medium density and orbital velocity.

Panel (b) of Figure~\ref{fig:f15} shows the efficiency of gas stripping in YZiCS galaxies depending on the magnitude of the ram pressure. From the analytic description of ram pressure stripping by \cite{Gunn&Gott_1972}, the amount of stripped gas depends on the ram pressure intensity and gravitational restoring force of a galaxy. In this paper, the stripping efficiency is measured at every snapshot as follows: 
\begin{equation} \label{eq:def_gas_loss}
\begin{aligned}
& \textrm{Stripping efficiency} = \frac{\textrm{Total gas loss} - \textrm{SF consumption}}{m_{\rm 0}}\\
& \textrm{Total gas loss} = m_{\rm 0}-m_{\rm 1} \\
& \textrm{SF consumption} = \Sigma (\textrm{SFR} \times \Delta t)\,, \\
\end{aligned}
\end{equation}
where $m_{\rm 0}$ is the gas mass of a galaxy at the previous snapshot, $m_{\rm 1}$ is the gas mass at the following snapshot, and $\Delta t$ is the time interval between two snapshots, roughly 70$\,\rm Myr$. The amount of gas consumption through star formation is measured by counting the number of star particles that are created between snapshots. When estimating the amount of gas stripped from the galaxy, we assume that all the decrease in the gas content, other than the consumption through star formation, is solely associated with this channel, because we cannot directly separate out gas elements escaping from the system in the AMR technique used in {\sc ramses}. Star particles in our prescription expel 10$\%$ of their mass into gas 10$\,\rm Myr$ after their formation, mimicking the mass loss through Type II SN explosion, but we do not consider the supply of gas through this channel for computing gas stripping efficiency. For this reason, our stripping efficiency is a lower limit.

In isolated environments or the outskirts of halos where the ram pressure is weak ($P_{\rm ram} / k_{\rm B} < 10^{4}$ $\rm cm^{-3}\,\rm K$), the stripping efficiency is effectively 0 (see panel (b) of Figure~\ref{fig:f15}). In this case, the gas consumption through star formation dominates the mass variation in the gas reservoir, and even if there may be some amount of gas stripped out from a galaxy, the effect is easily canceled out by the gas replenishment. However, for galaxies undergoing stronger ram pressure ($P_{\rm ram} / k_{\rm B} > 10^{4}$ $\rm cm^{-3}\,\rm K$), we find that the efficiency of gas stripping increases with increasing ram pressure intensity.

At the same time, at a given value of ram pressure, the stripping efficiency depends on the stellar mass of a galaxy. The blue boxes in Figure~\ref{fig:f15} show low mass galaxies ($< 10^{10}\, \rm M_{\rm \odot}$) with weak gravitational restoring force and the orange boxes represent massive galaxies ($> 10^{10}\, \rm M_{\rm \odot}$). We find that the lower mass galaxies lose more of their gas at given ram pressure. These results agree with the findings in Section \ref{s4}, that low mass satellite galaxies associated with massive group halos have a higher chance of being gas poor.

Figure~\ref{fig:f16} shows the star and gas distribution of six YZiCS galaxies with different ram pressure intensities selected from different redshift ranges. The galaxies in the upper row are experiencing gas stripping within cluster halo, while the bottom row shows the galaxies that are undergoing group pre-processing outside clusters. The yellow contour shows the gas column density distribution separated by 0.4 dex in log scale. In all panels, even in the panels showing galaxies in group-size halos, the stripped gas tails extend along the ram pressure wind-blowing direction (cyan arrow). Despite the severe deformation of the gas morphology, the stellar distributions remain unaffected. These features are direct evidence of the ram pressure stripping often observed in cluster galaxies (\citealt{Abramson_2011}; \citealt{Chung_2007}; \citealt{Chung_2009}).

In Section \ref{s5.1}, we discussed the efficiency of ram pressure stripping by comparing the magnitude of the ram pressure and the mean thermal pressure of galactic gas ($\langle P_{\rm thermal} \rangle$) that correlates with gravitational restoring force, in each galaxy. When considering the spatial variation of thermal pressure of galactic gas within a galaxy, even the galaxies undergoing strong ram pressure usually have gas cells whose thermal pressure is larger than the ram pressure in action in their central region (the green contour). These cells ($P_{\rm thermal} > P_{\rm ram}$) can be considered ``safe zone'' against ram pressure stripping. 
%The others ($P_{\rm thermal} > P_{\rm ram}$) are much easier to be pushed and stripped.

We find in our simulation that ram pressure stripping is at work in groups as well as clusters. However, discovery of the galaxies with disturbed gas morphology has been rare in group environments, which might appear in contradiction to our result. We can understand this by inspecting the magnitudes of the ram pressure that typical galaxies in groups and clusters experience. Panel (a) of Figure~\ref{fig:f15} shows that the difference in ram pressure between the satellites in the central regions (red diamonds) of groups (e.g., $12.2 < \log \, M_{\rm host} / M_{\odot} < 13$) and clusters ($13.8 < \log \, M_{\rm host} / M_{\odot}$) is by a factor of 40 ($\log \, P_{\rm ram}  / k_B \sim 3.75 \, {\rm cm^{-3} \, K}$ for group satellites and $\sim 5.35$ for cluster satellites). Thus, ram pressure stripping in groups is much more difficult to detect at any specific moment of observation. In fact, the most disturbed group satellite sample in Figure~\ref{fig:f16} (bottom left panel) shows an extremely rare case with a high ram pressure. Despite that, the cumulative amount of ram pressure stripping in groups is significant simply because it cumulates typically for more than a few gigayears before galaxies fall in a large cluster, whereas the typical cluster processing timescale is only about 1\,Gyr.

\subsection{Caveats}

The resolution of our simulation does not resolve detailed features of the Kelvin-Helmholtz instability at the interface between the galactic gas and the ICM, and thus it is likely to underestimate gas loss in satellite galaxies (e.g., \citealt{Nulsen_1982}; \citealt{Roediger_2005}; \citealt{Roediger_2007}). However, compared with the observational references introduced in Section \ref{s1} that galaxies keep forming stars for several gigayears after entering the clusters, YZiCS galaxies seem to lose too much of their cold gas and terminate their star formation too quickly even without continuous stripping. Here, we discuss possible origins for this disparity.

First, although our simulations adopt a spatial resolution ($\sim 0.76\,h^{-1} {\rm kpc}$) comparable to the current state-of-the-art cosmological simulations, such as Horizon-AGN \citep{Dubois2014}, Illustris \citep{Vogelsberger2014}, or Eagle \citep{Schaye_2015}, it is still not sufficient enough to resolve the detailed structure of the multi-phase ISM. For example, using the solar neighborhood conditions for star formation and radiation field, \citet{Kim_2013} showed that the total pressure in a cold and dense medium can be as high as $P_{\rm thermal}/k_B\sim 10^5\,{\rm cm^{-3} \, K}$ \citep[see also][for a gas-rich, dwarf galaxy environment]{Kimm_18}. The pressure can increase even further if infrared dark clouds ($n_{\rm H}\ga 10^4\,{\rm cm^{-3}}$) are concerned, suggesting that star-forming regions are likely to be resilient to the ram pressure stripping in group/cluster environments (i.e. $P_{\rm ram}/k_B\ga 10^{4-5}\,{\rm cm^{-3} \, K}$). Indeed, by observing 40 Virgo spiral galaxies, \cite{Kenney_1989} reported that cluster galaxies are not as much deficient in H$_{\rm 2}$ as in HI. Similarly, with spatially resolved observations, \cite{Vollmer_2012} discovered an increased value of H$_{\rm 2}$ to total gas fraction on the ram pressure windward side of the galactic gas disk. On the other hand, we find that the pressure of cold gas in our simulated galaxies is $P_{\rm thermal}/k_B \sim 10^4\,\rm cm^{-3}$, as the dense structures are artificially smoothed out due to finite resolution. In this regard, it is possible that the effect of ram pressure stripping is over-estimated and that star formation is overly quenched in our simulations.

The absence of magnetic fields is another issue. \citet{ryu00} and \citet{hennebelle13} demonstrated that the Kelvin-Helmholtz instability may be suppressed in small-scale filaments if strong magnetic tension is present. Since instabilities can accelerate the stripping process, we speculate that the neglect of magnetic fields in our simulations may also have resulted in too efficient gas stripping. At the same time, using a set of numerical simulations of a Milky-way-like disk, \citet{Tonnesen_2014} found that the gas initially pushed away from the mid-plane of the disk by the ICM wind can fall back more efficiently in the case with stronger magnetic tension. Interestingly, they find that the stripping rate in the simulations with different magnetic field strengths becomes eventually comparable \citep[c.f.][]{Ruszkowski_2014}, perhaps because strong magnetic pressure leads to a thicker disk, which in turn makes the low-density gas easier to be removed from the galaxy. However, their resolution ($159\,{\rm pc}$) may not be enough to resolve the disk scale-height, and more importantly, given that their simulations do not include radiative cooling, the thickness of the galactic disk may be overestimated \citep[e.g.][]{Kim_2013}, possibly artificially enhancing the stripping rate. Thus, high-resolution magneto-hydrodynamic simulations with radiative cooling will be needed to make a firm conclusion on the impact of magnetic fields on ram pressure stripping.

\section{Conclusion : Three channels of producing gas-poor galaxies} \label{s7}

\begin{figure*}
    \centering
    \includegraphics[width=\textwidth]{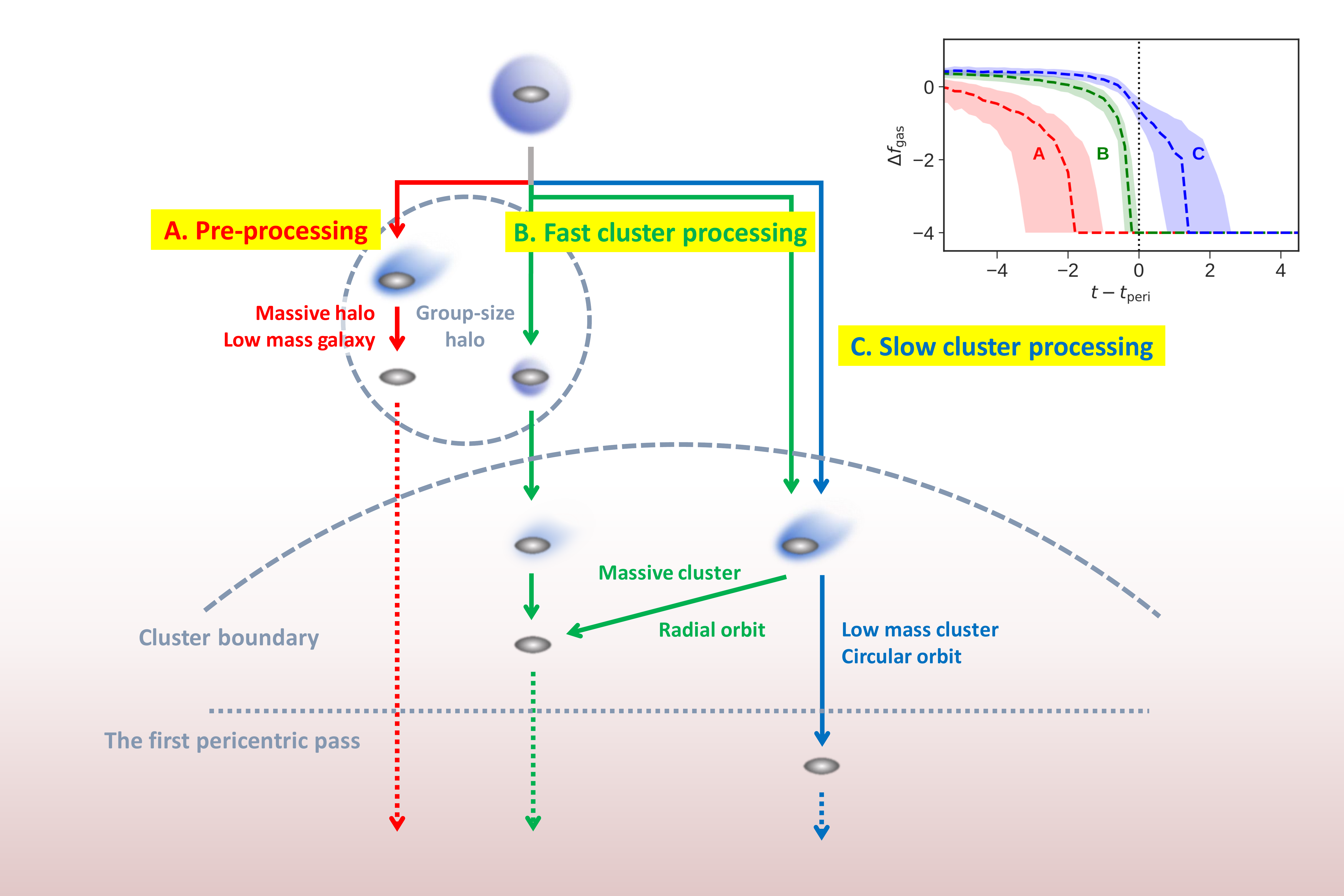}
    \caption{
    Illustration of the three channels that produce the gas-poor galaxies found in clusters.
    }  
    \label{fig:f17}
\end{figure*}

\begin{table*}
 \centering
 \begin{threeparttable}
 \caption{The relative importance of the three channels that produce gas-poor galaxies within the cluster virial radius.}
 \label{tab:t2}
 \begin{tabular}{ccccccc}
  \hline
  \parbox[t][][c]{0.6cm}{} &  & \parbox[t][][c]{2.4cm}{\centering Frac. Total sample} & \parbox[t]{2.4cm}{\centering Frac. MF weighted\\(16 clusters)} & \parbox[t]{2.4cm}{\centering Frac. MF weighted\\(13.5<$\log$ M<14)} & \parbox[t]{2.4cm}{\centering Frac. MF weighted\\(14<$\log$ M<14.5)} & \parbox[t]{2.5cm}{\centering Frac. MF weighted\\(14.5<$\log$ M<15)} \\ \hline
  A & Pre-processing & \parbox[t][][t]{2.4cm}{\centering33.7$\%$\\(647/1922)} & 21.7$\%$ & 17.4$\%$ & 31.9$\%$ & 35.9$\%$ \\ \hline
  B & Fast cluster processing & \parbox[t][][t]{2.4cm}{\centering42.7$\%$\\(820/1922)} & 33.8$\%$  & 29.7$\%$ & 43.5$\%$ & 49.7$\%$ \\ \hline
  C & Slow cluster processing & \parbox[t][][t]{2.4cm}{\centering23.7$\%$\\(455/1922)} & 44.5$\%$ & 52.9$\%$ & 24.6$\%$ & 14.3$\%$ \\ 
  \hline
 \end{tabular}
 
 \begin{tablenotes}
  \small
   \item Note. The first column shows the fractions obtained with a simple count of galaxies from the total sample. The rest of the columns show the fractions weighted with the analytic halo mass function (\citealt{Tinker_2008}), in order to avoid sampling bias. The second column is the weighted fraction for 16 clusters in total, and columns (3), (4), and (5) correspond to clusters in different mass ranges.
 \end{tablenotes}  
 \end{threeparttable}
\end{table*}

Throughout this paper, we have discussed how galaxies lose their gas and become gas poor in and outside the clusters by monitoring the gas content of 3818 galaxies from our 16 YZiCS clusters. We will now combine the discussions of the previous sections to describe the overall picture of the channels producing gas-poor galaxies now found inside clusters. 

Among the variety of gas depletion histories of our galaxies, we classify the cluster galaxies that are currently gas deficient into three groups based on where and when their gas depletion occurred. First, ``pre-processed'' galaxies are those that became gas poor already in group-size halos before the cluster entry. Second, ``fast cluster-processed'' galaxies are quickly depleted in their cold gas inside clusters during the first infall. Third, the ``slow cluster-processing'' channel is for the galaxies that retain cold gas for several gigayears after they arrive at the clusters even beyond their first pericentric pass. 
%Interestingly and importantly, {\em each channel can be explained with a different group environmental effect before entering clusters}. 
We summarize our understanding on each channel below and provide a schematic diagram of the possible scenarios in Figure~\ref{fig:f16}.\\

%\setlength{\parindent}{0cm}
%\textbf{\textit{The pre-processing channel}}
%\setlength{\parindent}{0.125in}

%\begin{enumerate}[label={(\roman*)}]
    %\item 
    Firstly, ``pre-processing'' is efficient for the low mass satellite galaxies in massive halos. More massive group halos have a higher intragroup medium density and higher velocity dispersion among satellite galaxies, resulting in a stronger ram pressure acting on their satellites. Meanwhile, under the given intensity of ram pressure, low mass galaxies lose their gas more efficiently as their gravitational restoring force is small.
    %\item 
    The magnitude of ram pressure in group-size halos (> $10^{12} \rm M_{\rm \odot}$) is enough to induce cold gas stripping. Some group satellite galaxies undergoing ram pressure stripping show disturbed gas morphology extending toward the opposite direction of the motion.
    %\item 
    In our simulation, once satellite galaxies exhaust their gas (i.e., less than 10\% of normal galaxies), they are rarely replenished in cold gas in group environments.
    %, i.e., the mechanism that made cold gas disappear prevents the rejuvenation of the galaxy.
    %\item 
    More massive clusters are likely to accrete higher fractions of pre-processed galaxies since they usually develop in overdense regions that have higher chances of attracting massive groups.
    %\item 
    The typical pattern of the gas loss history of such galaxies is shown as `A' in the inset of Figure~\ref{fig:f16}.
%\end{enumerate}

%\setlength{\parindent}{0cm}
%\textbf{\textit{The fast cluster processing channel}}
%\setlength{\parindent}{0.125in}

%\begin{enumerate}[label={(\roman*)}]
    %\item 
    Secondly, ``fast cluster processing'' takes place mainly but not exclusively on the galaxies that arrive at a cluster with a small amount of gas. 
    %quickly lose their gas during the first infall. 
    %experience the most dramatic environmental effect as they pass through the cluster pericenter for the first time. 
    %\item The galaxies that arrive at a cluster with a small amount of gas quickly lose their gas during the first infall.
    %Whether or not a galaxy runs out of its gas through the first pericenter pass is dependent on the gas depletion timescale, as the first-crossing timescale (time between the cluster infall and reaching the first pericenter, $\gsim 1-2\Gyr$) is constant among our clusters.
    They are usually those that have been a satellite galaxy before the cluster entry.
    %in the past halo history arrive at the cluster with less gas and hence more easily become gas poor during the first infall.
    %Satellite galaxies that have undergone environmental effects in group-size halos prior to the cluster infall usually arrive at the clusters with a decreased amount of gas compared to the central galaxies. These galaxies generally become gas poor before the first pericentric pass, almost regardless of their stellar mass, orbital circularity, and infalling cluster mass.
    In this case, other details such as stellar mass, orbital shape, or cluster mass are only of secondary importance.
    %\item 
    In the case of galaxies arriving at a cluster with a significant amount of gas, if they take the radial orbit toward the cluster center and/or fall into an extremely massive cluster ($> 10^{14.5}\,\rm M_{\rm \odot}$), they may run out of their gas through the first pericenteric pass, too.
    %\item 
    The typical pattern of the gas loss history of such galaxies is shown as `B' in the inset of Figure~\ref{fig:f16}.
%\end{enumerate}

%\setlength{\parindent}{0cm}
%\textbf{\textit{The slow cluster processing channel}}
%\setlength{\parindent}{0.125in}

%\begin{enumerate}[label={(\roman*)}]
    %\item 
    Lastly, there is ``slow cluster processing''. Galaxies that have always been a central galaxy are generally gas-rich at the time they enter clusters.
    %\item 
    When a gas-rich galaxy arrives at a low mass cluster ($< 10^{14}\,\rm M_{\rm \odot}$) and takes a circular orbit that does not approach the cluster center closely enough, the galaxy loses its gas very slowly.
    %\item 
    The typical pattern of the gas loss history of such galaxies is shown as `C' in the inset of Figure~\ref{fig:f17}.
%\end{enumerate}

The relative importance of these channels is presented in Table \ref{tab:t2}. At redshift 0, 1922 of our YZiCS galaxies reside within 1$\,R_{\rm 200}$ of each cluster and are gas deficient. Tracing back their gas content, we find that 33.7$\%$ and 42.7$\%$ take the pre-processing and fast cluster processing channels, respectively, and the rest, 23.7$\%$, are depleted in gas after their first pericentric pass (see the first column).

However, taking the numbers from the total galaxy sample blindly can lead us to a biased conclusion as more massive clusters own a larger number of satellites and therefore contribute more to the sample. Also, note that the YZiCS clusters are arbitrarily selected halos from a 200$\,\rm Mpc/h$-size-cube. Therefore, to avoid the sampling bias, we weight our 16 clusters using the analytic halo mass function (\citealt{Tinker_2008}) and then calculate the weighted average (see columns (2), (3), (4), and (5) in Table \ref{tab:t2}). The contribution of the slow cluster processing has notably been increased mainly by the consideration of low mass halos (second column). Separating our clusters into low (13.5 < $\log \, M _{\rm cluster}/\rm M_{\rm \odot}$ < 14, third column), intermediate (14 < $\log \, M _{\rm cluster}/\rm M_{\rm \odot}$ < 14.5, fourth column), and high (14.5 < $\log \, M _{\rm cluster}/\rm M_{\rm \odot}$ < 15, fifth column) mass ranges, we discover that the contribution of each channel varies with the mass of the cluster. In low mass clusters where the cluster environmental effect is relatively weak, the majority of their satellites contain gas by the time they pass their first pericenter (slow cluster processing). However, in massive clusters, the contribution of the slow cluster processing diminishes and most of the currently gas-poor cluster galaxies were produced during their first approach to the cluster center (fast cluster processing). With increasing cluster halo mass, we also find that a higher fraction of satellite galaxies was already gas deficient before they arrive at the current cluster (pre-processing).

Using the 16 clusters from YZiCS, we traced the gas depletion history of individual galaxies and investigated possible channels that produce gas-poor galaxies that are current members of clusters. As often regarded in previous studies, clusters are the harshest environments in the Universe, and the majority of their gas-poor members were produced after those galaxies met current clusters. However, a non-negligible fraction of galaxies was gas deficient, even before the cluster infall, i.e., pre-processed.

In previous studies, pre-processing and the cluster environmental effect have been studied separately, even though these two processes always act consecutively in the evolution of a galaxy falling into a cluster. According to our simulations, galaxies have a variety of gas fractions when they first arrive at clusters. We find that the previous environmental history is an important factor that determines the gas fraction. As a consequence, {\em whether a galaxy goes through pre-processing, fast cluster processing or slow cluster processing channels predominantly depends on the degree of the group environmental effect that the galaxy went through before entering the cluster}. 
%For example, even if group-influenced satellite galaxies do not completely run out of their gas outside of the clusters, they are low in gas by the time they arrive at the clusters, and therefore, lose their gas faster than ordinary gas-rich galaxies. Our study highlights the importance of the past history of galaxies, especially in group halos, before joining the current cluster in understanding the role of the environmental effect and the excess of passive galaxies in clusters.

\section*{Acknowledgements}
We thank Hyein Yoon and Sree Oh for the constructive feedback. S.K.Y. acknowledges support from the Korean National Research Foundation (NRF-2017R1A2A05001116). This study was performed under the umbrella of the joint collaboration between Yonsei University Observatory and the Korean Astronomy and Space Science Institute. The supercomputing time for numerical simulation was kindly provided by KISTI (KSC-2014-G2-003), and large data transfer was supported by KREONET, which is managed and operated by KISTI.

\bibliographystyle{apj}

\bibliography{mybib}

\end{document}